\documentclass[aps,prc,twocolumn,superscriptaddress,nofootinbib,longbibliography,floatfix,10pt]{revtex4-1}

\usepackage{graphicx}
\usepackage{dcolumn}
\usepackage{bm}
\usepackage{morefloats}
\usepackage{multirow}
\usepackage{amssymb}
\usepackage{amsmath}
\usepackage{xcolor}
\usepackage{fix-cm}
\usepackage{mathptmx} 
\usepackage[T1]{fontenc}
\usepackage[colorlinks,allcolors=blue]{hyperref}
\makeatletter
\def\NAT@def@citea{\def\@citea{\NAT@separator}}
\makeatother

\begin{document}

\title{Multi-Nucleon Transfer in Central Collisions of $^{238}$U + $^{238}$U}
\author{S. Ayik}\email{ayik@tntech.edu}
\affiliation{Physics Department, Tennessee Technological University, Cookeville, TN 38505, USA}
\author{B. Yilmaz}
\affiliation{Physics Department, Faculty of Sciences, Ankara University, 06100 Ankara, Turkey}
\author{O. Yilmaz}
\affiliation{Physics Department, Middle East Technical University, 06800 Ankara, Turkey}
\author{A. S. Umar}
\affiliation{Department of Physics and Astronomy, Vanderbilt University, Nashville, TN 37235, USA}
\author{G. Turan}
\affiliation{Physics Department, Middle East Technical University, 06800 Ankara, Turkey}

\date{\today}

\begin{abstract}
Quantal diffusion mechanism of nucleon exchange is studied in the central collisions of $^{238}$U +
$^{238}$U in the framework of the stochastic mean-field (SMF) approach. For bombarding energies
considered in this work, the di-nuclear structure is maintained during the collision. Hence, it is
possible to describe nucleon exchange as a diffusion process for mass and charge asymmetry. Quantal
neutron and proton diffusion coefficients, including memory effects, are extracted from the
SMF approach and the primary fragment distributions are calculated.
\end{abstract}


\maketitle

\section{Introduction}
Recently, much work has been done to investigate the multi-nucleon transfer processes in heavy-ion
collisions near barrier energies. For this purpose, the quasi-fission reaction of heavy-ions
provides an important tool. The colliding ions are attached together for a long time, but separate
without going through compound nucleus formation. During the long contact times many nucleon
exchanges take place between projectile and target nuclei. A number of models was developed for a
description of the reaction mechanism in the multi-nucleon transfer process in quasi-fission
reactions~\cite{adamian2003,zagrebaev2007,aritomo2009,zhao2016}.
Within the last few years the time-dependent Hartree-Fock (TDHF) 
approach~\cite{negele1982,nakatsukasa2016,simenel2012} has been
utilized for studying the dynamics of
quasifission~\cite{simenel2012,golabek2009,kedziora2010,simenel2012b,wakhle2014,oberacker2014,hammerton2015,umar2015c,umar2015a,sekizawa2016,umar2016}
and scission dynamics~\cite{simenel2014a,scamps2015a,simenel2016a,goddard2015,goddard2016,bulgac2016}.
Such calculations are now numerically feasible to perform on a
3D Cartesian grid without any symmetry restrictions
and with much more accurate numerical methods\,\cite{umar2006c,maruhn2014,schuetrumpf2016}.

The mean-field description of reactions using TDHF provides the mean values of the proton and neutron drift.
It is also possible to compute the probability to form a fragment with a given number of
nucleons~\cite{koonin1977,simenel2010,sekizawa2013,scamps2013a,sekizawa2015,sekizawa2015b}, 
but the resulting fragment mass and charge 
distributions are often underestimated in dissipative collisions~\cite{dasso1979,simenel2011}.
Much effort has been done to improve the standard mean-field approximation by incorporating the fluctuation mechanism
into the description. At low energies, the mean-field fluctuations make the dominant contribution to the fluctuation
mechanism of the collective motion.
Various extensions have been developed to study the fluctuations of one-body observables.
These include the TDRPA approach of Balian and V\'en\'eroni~\cite{balian1992}, the time-dependent generator coordinate 
method~\cite{goutte2005}, or the stochastic mean-field (SMF) method~\cite{ayik2008}.
The effects of two-body dissipation on reactions of heavy systems using the TDDM~\cite{tohyama1985,tohyama2002a},
approach have also been recently reported~\cite{assie2009,tohyama2016}.
Here we discuss some recent results using the SMF method~\cite{ayik2015a}.

In the SMF approach dynamical description is extended beyond the standard
approximation by incorporating the mean-field fluctuations into the description~\cite{ayik2008}. In
a number of studies, it has been demonstrated that the SMF approach is a good remedy for this shortcoming of the
mean-field approach and improves the description of the collisions dynamics by including
fluctuation mechanism of the collective motion~\cite{lacroix2014,yilmaz2014,ayik2015a,tanimura2017}.
Most applications have been carried out in collisions where a di-nuclear structure is maintained. In
this case, it is possible to define macroscopic variables with the help of the window dynamics. The
SMF approach gives rise to a Langevin description for the evolution of macroscopic variables
\cite{gardiner1991,weiss1999} and provides a microscopic basis to calculate transport coefficients
for the macroscopic variables. In most application, this approach has been applied to the nucleon
diffusion mechanism in the semi-classical limit and by ignoring the memory effects. In a recent work,
we were able to deduce the quantal diffusion coefficients for nucleon
exchange in the central collisions of heavy-ions~\cite{ayik2016} from the SMF approach. 
The quantal transport coefficients
include the effect of shell structure, take into account the full geometry of the collision process,
and incorporate the effect of Pauli blocking exactly. We applied the quantal diffusion
approach and carried out calculations for the variance of neutron and proton distributions of the
outgoing fragments in the central collisions of several symmetric heavy-ion systems at bombarding
energies slightly below the fusion barriers~\cite{ayik2016}. In this work we carry out quantal nucleon diffusion
calculations and determine the primary fragment mass and charge distributions in the central
collisions of $^{238}$U + $^{238}$U system in side-side and tip-tip configurations. Since the
presented calculations do not involve any fitting parameters, the results may provide a useful
guidance for the experimental investigations of heavy neutron rich isotopes originating from these reactions.

In section 2, we present a brief description of the quantal nucleon diffusion mechanism based on the
SMF approach. In section 3, we present a brief discussion of quantal neutron and proton diffusion
coefficients. The result of calculations is reported in section 4, and conclusions are given in
section 5.

\section{Nucleon diffusion description}
\begin{figure}[!hpt]
\vspace{0.2cm}
\includegraphics[width=8.5cm]{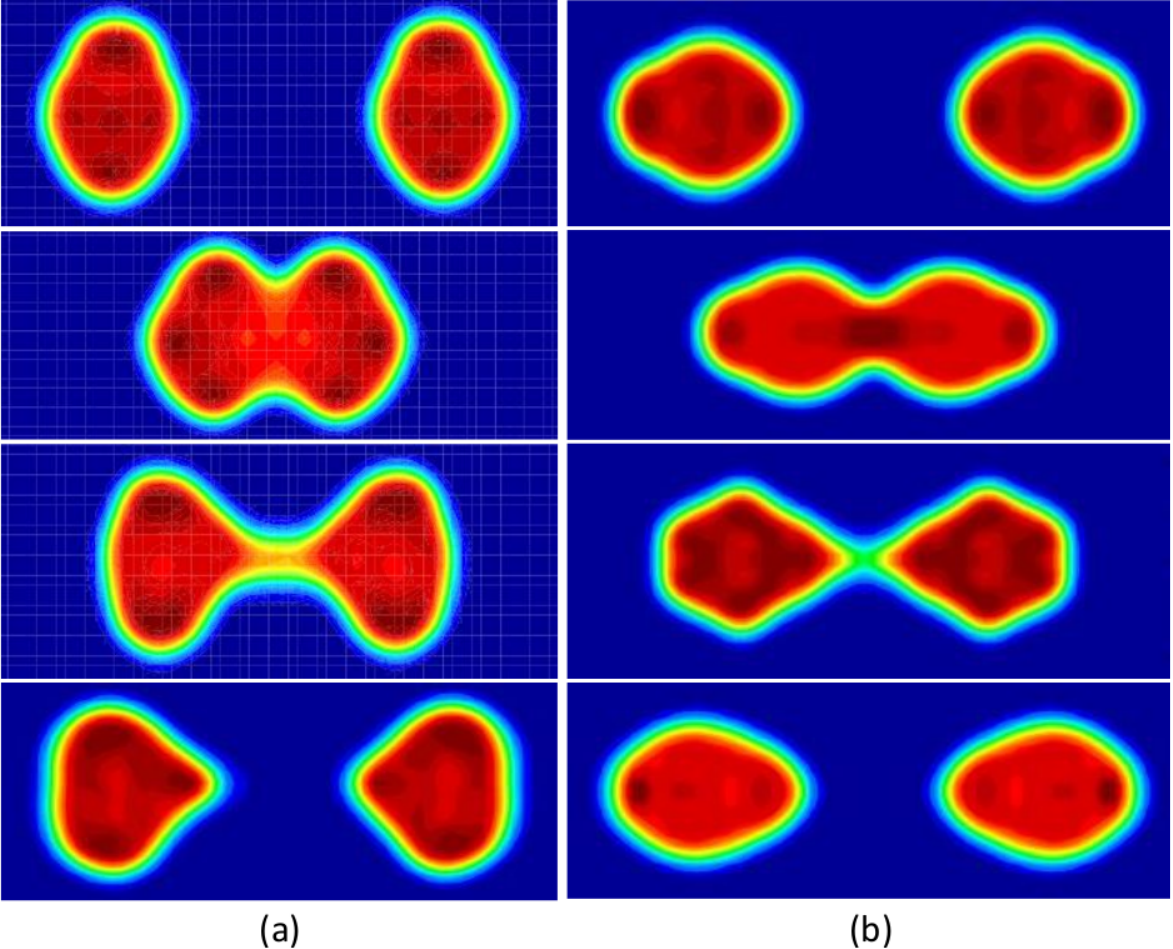}
\vspace{0.2cm}
\caption{(Color online) Density profiles in the reaction plane in the central collisions of
$^{238}$U + $^{238}$U (a) side-side collision with energy $E_{c.m.} =900$~MeV from top to bottom at
times t= 0, 400, 800, and 950 fm/c, and (b) tip-tip collision with energy $E_{c.m.} =1050$~MeV from
top to bottom at times t= 0, 200, 700, and 800 fm/c, respectively, obtained in TDHF calculations.
\label{fig1}}
\end{figure}

In heavy-ion collisions when the system maintains a binary structure, the reaction evolves mainly
due to nucleon exchange through the window between the projectile-like and target-like partners. It
is possible to analyze nucleon exchange mechanism by employing nucleon diffusion concept based on
the SMF approach. In the SMF approach, the standard mean-field description is extended by
incorporating the mean-field fluctuations in terms of generating an ensemble of events according to
quantal and thermal fluctuations in the initial state. Instead of following a single path, in the
SMF approach dynamical evolution is determined by an ensemble of Slater determinants. The initial
conditions of the single-particle density matrices associated with the ensemble Slater determinants
are specified in terms of the quantal and thermal fluctuations of the initial state. For a detailed
description of the SMF approach, we refer to~\cite{ayik2008,lacroix2014,yilmaz2014,ayik2015a}. In
extracting transport coefficients for nucleon exchange, we take the proton and neutron numbers of
projectile-like fragments $Z_{1}^{\lambda }$, $N_{1}^{\lambda}$ as independent variables, where
$\lambda$ indicates the event label. We can define the proton and neutron numbers of the
projectile-like fragments in each event by integrating over the nucleon density on the projectile
side of the window. In the central collisions of symmetric systems, the window is perpendicular to
the collision direction taken as the $x$-axis and the position of the window is fixed at the origin
of the center of mass frame according to the mean-field description of the TDHF. The proton and
neutron numbers of the projectile-like fragments in each events are defined as,
\begin{align} \label{eq1}
\left(\begin{array}{c} {Z_{1}^{\lambda } (t)} \\ {N_{1}^{\lambda } (t)} \end{array}\right)=\int
d^{3} r\theta (x-x_{0} ) \left(\begin{array}{c}{\rho_{p}^{\lambda}(\vec{r},t)} \\ {\rho
_{n}^{\lambda } (\vec{r},t)} \end{array}\right).
\end{align}
Here, $x_{0} =0$ denotes average position of the window plane taken as the origin of the center of
mass frame and $\rho _{p}^{\lambda}(\vec{r},t)$ and $\rho_{n}^{\lambda}(\vec{r},t)$ are the local
densities of protons and neutrons. Fig.~\ref{fig1} shows the evolution of the average density profile in the
side-side and tip-tip collisions of $^{238}$U + $^{238}$U nuclei at bombarding energies
E$_{c.m.}=900$~MeV and E$_{c.m.} =1050$~MeV, respectively. In the calculation of this figure and
in the calculations presented in the rest of the article, we employ the TDHF code developed by 
Umar et al.~\cite{umar1991a,umar2006c} using the SLy4d Skyrme functional~\cite{kim1997}.

In the collision of symmetric systems, location of the window plane remains stationary, and on
the average, there is no net nucleon transfer between projectile and target nuclei. According to the
SMF approach, the proton and neutron numbers of the projectile-like fragment follows a stochastic
evolution according to the Langevin equations,
\begin{align} \label{eq2}
\frac{d}{dt}\left(\begin{array}{c} {Z_{1}^{\lambda } (t)} \\ {N_{1}^{\lambda } (t)}
\end{array}\right) &= \int d^{3} rg(x)\left(\begin{array}{c} {J_{p}^{\lambda } (\vec{r},t)} \\
{J_{n}^{\lambda } (\vec{r},t)} \end{array}\right) \nonumber\\ &=\left(\begin{array}{c}
{v_{p}^{\lambda } (t)} \\ {v_{n}^{\lambda } (t)} \end{array}\right).
\end{align}
In this expression, in place of the delta function $\delta (x)$ we introduce a Gaussian smoothing
function $g(x)$ for convenience,
\begin{align} \label{eq3}
g(x)=\frac{1}{\sqrt{2\pi\kappa^2}}\exp \left(-\frac{x^2}{2\kappa^2}\right),
\end{align}
which approaches the delta function $\delta(x)$ in the limit $\kappa \to 0$. For the smoothing
parameter, we take the value $\kappa=1$ fm. This value is in the order of lattice spacing of the
numerical calculations performed in this work. The right hand side of Eq.~\eqref{eq2} denotes the
proton $v_{p}^{\lambda }(t)$ and neutron $v_{n}^{\lambda}(t)$ drift coefficients in the event
$\lambda$, which are determined by the proton and the neutron current densities,
$J_{p}^{\lambda}(\vec{r},t)$ and $J_{n}^{\lambda}(\vec{r},t)$, through the window in that event. In the
SMF approach, the fluctuating proton and neutron currents densities in the collision direction are
determined to be,
\begin{align} \label{eq4}
J_{\alpha }^{\lambda } (\vec{r},t)=\frac{\hbar }{m} \sum _{ij\in \alpha }Im\left(\Phi _{j}^{*}
(\vec{r},t;\lambda )\nabla _{x} \Phi _{i} (\vec{r},t;\lambda )\right) \rho _{ji}^{\lambda }.
\end{align}
Here, and in the rest of the paper, we use the label $\alpha=p,n$ for the proton and neutron states.
In the description of the SMF approach, the elements of density matrices $\rho _{ji}^{\lambda }$ are
taken as uncorrelated Gaussian numbers. The mean values of the elements of density matrices are given
by $\overline{\rho }_{ji}^{\lambda}=\delta _{ji} n_{j}$ and the second moments of fluctuating parts
are determined by
\begin{align} \label{eq5}
\overline{\delta \rho _{ji}^{\lambda } \delta \rho _{i'j'}^{\lambda } }=\frac{1}{2} \delta _{ii'}
\delta _{jj'} \left[n_{i} (1-n_{j} )+n_{j} (1-n_{i} )\right],
\end{align}
where $n_{j}$ are the average occupation numbers of the single-particle states.

For small amplitude fluctuations, by taking the ensemble averaging, we obtain the usual mean-field
result given by the TDHF equations,
\begin{align} \label{eq6}
\frac{d}{dt} \left(\begin{array}{c} {Z_{1} (t)} \\ {N_{1} (t)} \end{array}\right)&=\int d^{3}
rg(x)\left(\begin{array}{c} {J_{p} (\vec{r},t)} \\ {J_{n} (\vec{r},t)} \end{array}\right)
\nonumber\\ &=\left(\begin{array}{c} {v_{p} (t)} \\ {v_{n} (t)} \end{array}\right).
\end{align}
Here, $Z_{1} =\overline{Z}_{1}^{\lambda}$, $N_{1} =\overline{N}_{1}^{\lambda}$,
$J_{\alpha}(\vec{r})=\overline{J}_{\alpha }^{\lambda}(\vec{r})$ and $v_{\alpha}
=\overline{v}_{\alpha}^{\lambda}$ indicate the mean values of the proton and neutron numbers of
projectile-like fragments, proton and neutron current densities, and proton and neutron drift
coefficients, which are average values taken over the ensemble single-particle densities. Mean
values of the current densities of protons and neutrons along the collision direction are given by,
\begin{align} \label{eq7}
J_{\alpha}(\vec{r},t)=\frac{\hbar}{m}\sum_{h\in \alpha }\text{Im}\left(\Phi _{h}^{*} (\vec{r},t)\nabla _{x} \Phi _{h} (\vec{r},t)\right),
\end{align}
where the summation $h$ runs over the occupied states originating both from the projectile and the
target nuclei. Drift coefficients $v_{p}^{\lambda}(t)$ and $v_{n}^{\lambda}(t)$ fluctuate from event
to event due to stochastic elements of the initial density matrix $\rho_{ji}^{\lambda}$ and also due
to the different sets of the wave functions in different events. As a result, there are two sources
for fluctuations of the nucleon current: (i) fluctuations those arise from the state dependence of
the drift coefficients, which may be approximately represented in terms of fluctuations of proton
and neutron partition of the di-nuclear system, and (ii) the explicit fluctuations $\delta
v_{p}^{\lambda}(t)$ and $\delta v_{n}^{\lambda}(t)$ which arise from the stochastic part of proton
and neutron currents. For small amplitude fluctuations, we can linearize the Langevin 
Eq.~\eqref{eq2} around the mean evolution to obtain,
\begin{align} \label{eq8}
\frac{d}{dt} \left(\begin{array}{c} {\delta Z_{1}^{\lambda } (t)} \\ {\delta N_{1}^{\lambda } (t)}
\end{array}\right)&=\left(\begin{array}{c} {\frac{\partial v_{p} }{\partial Z_{1} }
\left(Z_{1}^{\lambda } -Z_{1} \right)+\frac{\partial v_{p} }{\partial N_{1} } \left(N_{1}^{\lambda }
-N_{1} \right)} \\ {\frac{\partial v_{n} }{\partial Z_{1} } \left(Z^{\lambda } -Z_{1}
\right)+\frac{\partial v_{n} }{\partial N_{1} } \left(N_{1}^{\lambda } -N_{1} \right)}
\end{array}\right)\nonumber\\ &\quad+\left(\begin{array}{c} {\delta v_{p}^{\lambda } (t)} \\ {\delta
v_{n}^{\lambda } (t)} \end{array}\right).
\end{align}
The variances and the co-variance of neutron and proton distribution of projectile fragments are
defined as $\sigma _{NN}^{2}(t)=\overline{\left(N_{1}^{\lambda }-N_{1}\right)^{2} }$, $\sigma
_{ZZ}^{2}(t)=\overline{\left(Z_{1}^{\lambda }-Z_{1}\right)^{2} }$, and $\sigma
_{NZ}^{2}(t)=\overline{\left(N_{1}^{\lambda}-N_{1}\right)\left(Z_{1}^{\lambda}-Z_{1} \right)}$.
Multiplying both side of Eq.~\eqref{eq8} by $N_{1}^{\lambda }-N_{1}$ and $Z_{1}^{\lambda}-Z_{1}$,
and taking the ensemble average, it is possible to obtain set of coupled differential equations for
the co-variances~\cite{schroder1981,merchant1981}. These differential equations are given by,
\begin{align} \label{eq9}
\frac{\partial }{\partial t} \sigma _{NN}^{2} &=2\frac{\partial v_{n} }{\partial N_{1} } \sigma
_{NN}^{2} +2\frac{\partial v_{n} }{\partial Z_{1} } \sigma _{NZ}^{2} +2D_{NN},\\
\label{eq10}
\frac{\partial }{\partial t} \sigma _{ZZ}^{2} &=2\frac{\partial v_{p} }{\partial Z_{1} } \sigma
_{ZZ}^{2} +2\frac{\partial v_{p} }{\partial N_{1} } \sigma _{NZ}^{2} +2D_{ZZ},\\
\label{eq11}
\frac{\partial }{\partial t} \sigma _{NZ}^{2} &=\frac{\partial v_{p} }{\partial N_{1} } \sigma
_{NN}^{2} +\frac{\partial v_{n} }{\partial Z_{1} } \sigma _{ZZ}^{2}\nonumber\\ &\quad+\sigma
_{NZ}^{2} \left(\frac{\partial v_{p} }{\partial Z_{1} } +\frac{\partial v_{n} }{\partial N_{1} }
\right).
\end{align}
Here, $D_{NN}$ and $D_{ZZ}$ indicate the diffusion coefficients of proton and neutron exchanges. In
order to determine the co-variances in addition to the diffusion coefficients, we need to know
derivatives of drift coefficients with respect to the proton and neutron numbers. These derivatives
are evaluated at the mean values of the neutron and proton numbers. In symmetric collisions, mean
values of the drift coefficients are zero, but in general, their slopes at the zero mean values do
not vanish.

It is well know that the Langevin description is equivalent to the Fokker-Planck description of the
probability distribution function $P(N,Z,t)$ primary fragments as a function of the neutron and
proton numbers~\cite{risken1996}. When fluctuating drift coefficients are linear functions of the
fluctuating proton and neutron numbers, the probability distribution of the project-like or the
target-like fragments are specified by a correlated Gaussian function,
\begin{align} \label{eq12}
P(N,Z,t)=\frac{1}{2\pi \sigma _{NN} \sigma _{ZZ} \sqrt{1-\rho ^{2} } } \exp \left(-C\right).
\end{align}
Here, the exponent $C$ is given by
\begin{align} \label{eq13}
C=\frac{1}{2\left(1-\rho ^{2} \right)} &\left[\left(\frac{Z-\overline{Z}}{\sigma _{ZZ} }\right)^{2}+\left(\frac{N-\overline{N}}{\sigma _{NN}}\right)^{2}\right.\nonumber\\
&\left.\quad -2\rho \left(\frac{Z-\overline{Z}}{\sigma _{ZZ} } \right)\left(\frac{N-\overline{N}}{\sigma _{NN} } \right) \right],
\end{align}
where $\rho=\sigma_{NZ}^{2}/\sigma _{ZZ}\sigma_{NN}$ is the correlation coefficient. The mean values
$\overline{N}$, $\overline{Z}$ are the mean neutron and proton numbers of the target-like or
project-like fragments.

\section{Transport coefficients for nucleon exchange}

\subsection{Quantal diffusion coefficients}

The quantal expressions of the proton and neutron diffusion coefficients are determined by the
correlation function of the stochastic part of the drift coefficients according to
\cite{gardiner1991,weiss1999},
\begin{align} \label{eq14}
D_{\alpha \alpha } (t)=\int _{0}^{t}dt' \overline{\delta v_{\alpha }^{\lambda } (t)\delta v_{\alpha }^{\lambda } (t')}.
\end{align}
From Eq.~\eqref{eq4}, the stochastic parts of the drift coefficients are given by,
\begin{align} \label{eq15}
\delta v_{\alpha }^{\lambda } (t)=\frac{\hbar }{m} \sum _{ij\in \alpha }\int d^{3}
rg(x)\text{Im}\left(\Phi _{j}^{*} (\vec{r},t)\nabla _{x} \Phi _{i} (\vec{r},t)\right)  \delta \rho
_{ji}^{\lambda}.
\end{align}
In determining the stochastic part of the drift coefficients, we impose a physical constraint on the
summations of single-particle sates. The transitions among single particle states originating from
projectile or target nuclei do not contribute to nucleon exchange mechanism. Therefore, in this
expression, we restrict the summations as follows: when the summation $i\in T$ runs over the states
originating from target nucleus, the summation  $j\in P$ runs over the states originating from the
projectile, and vice versa.

Using the basic postulate of the SMF approach given by Eq.~\eqref{eq5}, it is possible to calculate
the correlation functions of the stochastic part of the drift coefficients, and hence we can
determine the quantal expression for the diffusion coefficients. The correlation function involves a
complete set of time-dependent particle and hole states. The standard solutions of TDHF give the
time-dependent wave functions of the occupied hole states. The solution of complete set of time-dependent
particle states requires a very large amount of effort. However, it is possible to eliminate the
complete set of particle states by employing closure relation with the help of a reasonable
approximation. We recognize that the time-dependent single-particle wave functions obtained from the
TDHF exhibit nearly a diabatic behavior~\cite{norenberg1981}. In other words, during
short time intervals the nodal structure of time-dependent wave functions do not change appreciably.
Most dramatic diabatic behavior of the time-dependent wave functions is apparent in the fission
dynamics. The Hartree-Fock solutions force the system to follow the diabatic path, which prevents
the system to break up into fragments.  As a result of these observations, we introduce, during
short time $\tau =t-t'$ evolutions in the order of the correlation time, a diabatic approximation
into the time-dependent wave functions by shifting the time backward (or forward) according to
\begin{align} \label{eq16}
\Phi _{a} (\vec{r},t')\approx \Phi _{a} (\vec{r}-\vec{u}\tau ,t),
\end{align}
where $\vec{u}$ denotes a suitable flow velocity of nucleons. Now, we can employ the closure relation,
\begin{align} \label{eq17}
\sum _{a}\Phi _{a}^{*} (\vec{r}_{1} ,t)\Phi _{a} (\vec{r}_{2} ,t') &\approx \sum _{a}\Phi _{a}^{*} (\vec{r}_{1} ,t)\Phi _{a} (\vec{r}_{2} -\vec{u}\tau ,t)\nonumber\\
&=\delta (\vec{r}_{1} -\vec{r}_{2} +\vec{u}\tau ),
\end{align}
where, summation $a$ runs over the complete set of states originating from target or projectile, and
the closure relation is valid for each set of the spin-isospin degrees of freedom. Carrying out an
algebraic manipulation, we find that the quantal expressions of the proton and neutron diffusion
coefficients are given by
\begin{align} \label{eq18}
D_{\alpha \alpha } (t)&=\int _{0}^{t}d\tau  G_{0} (\tau )\int d^{3} r \tilde{g}(x)\nonumber\\
&\quad\times\left[J_{\alpha }^{T} (\vec{r},t-\tau /2)+J_{\alpha }^{P} (\vec{r},t-\tau /2)\right]\nonumber\\
&\;\;-\int_{0}^{t}d\tau\text{Re}\left[\sum _{h'\in P,h\in T}A_{h'h}^{\alpha } (t)A_{h'h}^{*\alpha}(t-\tau)\right.\nonumber\\
&\qquad\qquad\qquad\left.+\!\!\!\!\sum_{h'\in T,h\in P}A_{h'h}^{\alpha}(t)A_{h'h}^{*\alpha}(t-\tau)\right],
\end{align}
where $\tilde{g}(x)=(1/\sqrt{\pi}\kappa)\exp[-(x/\kappa )^{2}]$. The quantity
$J_{\alpha}^{T}(\vec{r},t-\tau/2)$ represents the sum of
magnitude of the current densities due to hole wave functions originating from target
nuclei,
\begin{align} \label{eq19}
J_{\alpha }^{T} (\vec{r},t)=\frac{\hbar}{m}\sum_{h\in T}|\text{Im}\left(\Phi _{h}^{*} (\vec{r},t)\nabla _{x} \Phi _{h} (\vec{r},t)\right)| .
\end{align}
Here, the quantity $G_{0} (\tau )=[1/(\tau _{0} \sqrt{4\pi })]\exp[-(\tau /2\tau _{0})^{2}]$ denotes
the memory kernel with the memory time given by $\tau_{0}=\kappa /|u_{0}|$ with  $u_{0} =\langle
u_{h}\rangle$ as the average flow speed of hole states across the window. The quantity $J_{\alpha
}^{P}(\vec{r},t-\tau/2)$ associated with the projectile states is given by a similar expression. The
hole-hole matrix elements $A_{h'h}^{\alpha}(t)$ calculated with the wave functions originating from
projectile and target nuclei are given by,
\begin{align} \label{eq20}
A_{h'h}^{\alpha}(t)=\frac{\hbar}{2m}\int d^{3}r g(x)&\left[\Phi_{h'}^{*\alpha}(\vec{r},t)\nabla _{x}\Phi _{h}^{\alpha}(\vec{r},t)\right.\nonumber\\
&\left.-\Phi_{h}^{\alpha}(\vec{r},t)\nabla_{x}\Phi _{h'}^{*\alpha}(\vec{r},t)\right].
\end{align}
For a detailed derivation of quantal diffusion coefficients Eq.~\eqref{eq18} and definition of flow
velocities, we refer the reference~\cite{ayik2016}. There is a close analogy between the quantal
expression and the classical diffusion coefficient in a random walk 
problem~\cite{gardiner1991,weiss1999,randrup1979}. The first line in the quantal expression gives the sum of
the nucleon currents from the target-like fragment to the projectile-like fragment and from the projectile-like 
fragment to the target-like fragment, which is
integrated over the memory. This is analogous to the random walk problem, in which the diffusion
coefficient is given by the sum of the rate for the forward and backward steps. The second line in
the quantal diffusion expression stands for the Pauli blocking effects in nucleon transfer
mechanism, which does not have a classical counterpart. It is important to note that the quantal
diffusion coefficients are entirely determined in terms of the occupied single-particle wave
functions obtained from the TDHF solutions. The quantal diffusion coefficients contain the effects
of the shell structure, take into account full collision geometry and do not involve any free
parameters. In the collisions at the energies we considered, the average value of nucleon flow speed across
the window is $u_{0}\approx 0.05$c~\cite{ayik2016}, which gives a memory time around $\tau _{0}
=\kappa /u_{0} \approx 20$ fm/c. Since the memory time is much shorter than a typical interaction
time of collisions, $\tau_{0}<<500$ fm/c, the memory effect is not very effective in nucleon
exchange mechanism. Consequently, we can neglect the $\tau$ dependence in the current densities in
Eq.~\eqref{eq18}, carry out the $\tau$ integration over the memory kernel to give $\int
_{0}^{t}G_{0}(\tau)d\tau\approx1/2$. Because of the same reason, memory effect is not very effective
in the Pauli blocking terms as well, however in the calculations we keep the memory integrals in
these terms.

\subsection{Nucleon drift coefficients}

\begin{figure}[!hpt]
\includegraphics[width=8cm]{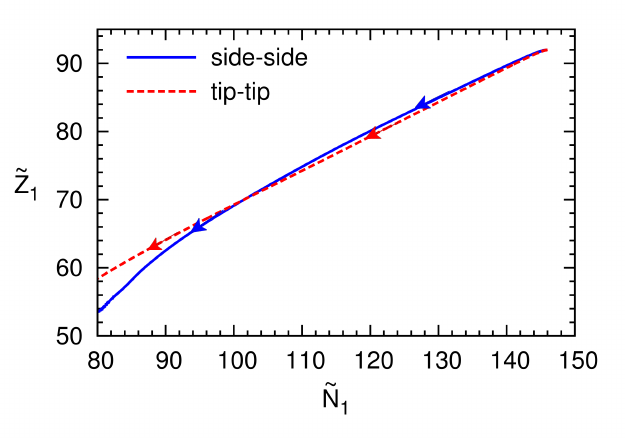}
\vspace{-0.5cm}
\caption{(Color online) Mean drift path of the projectile-like fragments in $N-Z$ plane in the
central collisions of $^{238}$U + $^{238}$U at side-side collision with energy $E_{c.m.} =900$~MeV
(solid line), and at tip-tip collision with energy $E_{c.m.} =1050$~MeV (dashed line), obtained with
one-sided flux in TDHF calculations.\label{fig2}}
\end{figure}

In order to solve co-variances from Eqs.~(\ref{eq9}-\ref{eq11}), in addition to the diffusion coefficients
$D_{ZZ}$ and $D_{NN}$, we need to know the rate of change of drift coefficients in the vicinity of
their mean values. According to the SMF approach, in order to calculate rates of the drift
coefficients, we should calculate neighboring events in the vicinity of the mean-field event. Here,
instead of such a detailed description, we employ the fluctuation-dissipation theorem, which
provides a general relation between the diffusion and drift coefficients in the transport mechanism
of the relevant collective variables as described in the phenomenological approaches
\cite{randrup1979}. Proton and neutron diffusions in the N-Z plane are driven in a correlated manner
by the potential energy surface of the di-nuclear system.  As a consequence of the symmetry energy, the
diffusion in direction perpendicular to the beta stability valley takes place rather rapidly leading
to a fast equilibration of the charge asymmetry, and diffusion continues rather slowly along the
beta-stability valley. Borrowing an idea from references~\cite{norenberg1981,merchant1982}, we
parameterize the $N_{1}$ and $Z_{1}$ dependence of the potential energy surface of the di-nuclear
system in terms of two parabolic forms,
\begin{align} \label{eq21}
U(N_{1} ,Z_{1} )=&\frac{1}{2} a\left(z\cos \theta -n\sin \theta \right)^{2}\nonumber\\
&+\frac{1}{2} b\left(z\sin \theta +n\cos \theta \right)^{2}.
\end{align}
Here, $z=Z_{0}-Z_{1}$, $n=N_{0}-N_{1}$ and $\theta$ denotes the angle between beta stability valley
and the $N$ axis in the $N-Z$ plane.  The quantities $N_{0}$ and $Z_{0}$ denote the equilibrium
values of the neutron and proton numbers, which are approximately determined by the average values
of the neutron and proton numbers of the projectile and target ions,
$N_{0}=\left(N_{P}+N_{T}\right)/2$ and $Z_{0}=\left(Z_{P} +Z_{T} \right)/2$. The first term in this
expression describes a strong driving force perpendicular to the beta stability valley, while the
second term describes a relative weak driving force toward symmetry along the valley. In symmetric
collisions, $N_{0}$ and $Z_{0}$ are equal to the initial neutron and proton numbers of the target or
projectile nuclei. Following from the fluctuation-dissipation theorem, it is possible to relate the
proton and neutron drift coefficients to the diffusion coefficients and the associated driving
forces, in terms of the Einstein relations as follows~\cite{norenberg1981,merchant1982},
\begin{align} \label{eq22}
\nu_n &=-\frac{D_{NN}}{T}\frac{\partial U}{\partial N_{1}} =+\frac{D_{NN}}{T}\frac{\partial U}{\partial n}\nonumber\\
&=D_{NN}\left[-\alpha \sin \theta \left(z\cos \theta -n\sin \theta \right)\right.\nonumber\\
&\qquad\qquad\left.+\beta \cos \theta \left(z\sin \theta +n\cos \theta \right)\right],
\end{align}
and
\begin{align} \label{eq23}
\nu_z &=-\frac{D_{ZZ}}{T}\frac{\partial U}{\partial Z_{1}} =+\frac{D_{ZZ}}{T}\frac{\partial U}{\partial z}\nonumber\\
&=D_{ZZ}\left[+\alpha \cos \theta \left(z\cos \theta -n\sin \theta \right)\right.\nonumber\\
&\qquad\qquad\left.+\beta \sin \theta \left(z\sin \theta +n\cos \theta \right)\right].
\end{align}
Here, the temperature $T$ is absorbed into coefficients $\alpha$ and $\beta$, consequently
temperature does not appear as a parameter in the description. In asymmetric collisions, it is
possible to determine $\alpha$ and $\beta$ by matching the mean values of neutron and proton drift
coefficients obtained from the TDHF solutions. In symmetric collisions, the mean value of drift
coefficients are zero and the mean values of neutron and proton numbers do not change and remain
equal to their initial values. Therefore it is not possible to determine the coefficients $\alpha$
and $\beta$ from the full TDHF solutions. However, we can determine these coefficients employing the
one-sided neutron and proton fluxes from projectile-like fragment to the target-like fragment or vice-versa. We
indicate neutron and proton numbers of one of the fragments as $\tilde{N}_{1}$ and $\tilde{Z}_{1}$.
Then, the neutron and proton numbers of this fragment monotonically decreases according to,
\begin{align} \label{eq24}
\frac{d}{dt} \left(\begin{array}{c} {\tilde{Z}_{1} (t)} \\ {\tilde{N}_{1} (t)}
\end{array}\right)&=\int d^{3} rg(x)\left(\begin{array}{c} {\tilde{J}_{p} (\vec{r},t)} \\
{\tilde{J}_{n} (\vec{r},t)} \end{array}\right)\nonumber\\ &=\left(\begin{array}{c} {\tilde{v}_{p}
(t)} \\ {\tilde{v}_{n} (t)} \end{array}\right).
\end{align}
Here, $\tilde{v}_{\alpha}(t)$ with $\alpha =n,p$ denotes the one-sided neutron and proton drift
coefficients towards the other fragment and the one-sided current density
$\tilde{J}_{\alpha}(\vec{r},t)$ is given by Eq.~\eqref{eq7} keeping only negative terms in the
summation over the hole states. The one-sided drift coefficients $\tilde{\nu}_{n}$ and $\tilde{\nu
}_{p}$ are related to the driving force with the similar expressions given by Eq.~\eqref{eq22} and
Eq.~\eqref{eq23}, except that $n$ and $z$ are replaced by $\tilde{n}=N_{0} -\tilde{N}_{1}$ and
$\tilde{z}=Z_{0} -\tilde{Z}_{1}$ and by including an overall sign change,
\begin{align} \label{eq25}
\tilde{\nu}_{n} =D_{NN}&\left[+\alpha \sin \theta \left(\tilde{z}\cos \theta -\tilde{n}\sin \theta \right)\right.\nonumber\\
&\;\left.-\beta \cos \theta \left(\tilde{z}\sin \theta +\tilde{n}\cos \theta \right)\right],
\end{align}
and
\begin{align} \label{eq26}
\tilde{\nu}_{z} =D_{ZZ}&\left[-\alpha \cos \theta \left(\tilde{z}\cos \theta -\tilde{n}\sin \theta \right)\right.\nonumber\\
&\;\left.-\beta \sin \theta \left(\tilde{z}\sin \theta +\tilde{n}\cos \theta \right)\right].
\end{align}
Fig.~\ref{fig2} shows the one-sided mean-drift paths of projectile-like fragments which are determined by
keeping the one-sided neutron and proton fluxes from projectile-like to the target-like fragments in
the side-side and tip-tip collisions of $^{238}$U + $^{238}$U.
Using this information, we can extract the angle $\theta$ and the magnitude of coefficients $\alpha
$ and $\beta$. We find that, the angle between the mean one-sided drift path and $N$-axis is about
$\theta \approx 30^\circ$ in both collision geometries. As a result of the quantal effects arising
mainly from the shell structure, we observe that the coefficients $\alpha$ and $\beta$ exhibit
fluctuations as a function of time. In the side-side collision, during the relevant time interval from
$200$ fm/c to $800$ fm/c, the average values of these coefficients are about $\alpha \approx 0.035$
and $\beta \approx 0.007$. In the tip-tip collision, during the relevant time interval from $200$ fm/c
to $700$ fm/c, the average values of these coefficients are about $\alpha \approx 0.039$ and $\beta
\approx 0.009$. These results are consistent with the potential energy surface of the liquid drop
picture. The potential energy surface in (N-Z) plane has a steeply rising parabolic shape in the 
perpendicular direction to the stability valley and has a shallow behavior along the
stability valley. Because of a simple
analytical structure, we can easily calculate derivatives of drift coefficients which are needed in
differential Eqs.~(\ref{eq9}-\ref{eq11}) for determining the co-variances.

\section{Primary fragment distributions}
\begin{figure}[!hpt]
\includegraphics[width=7.2cm]{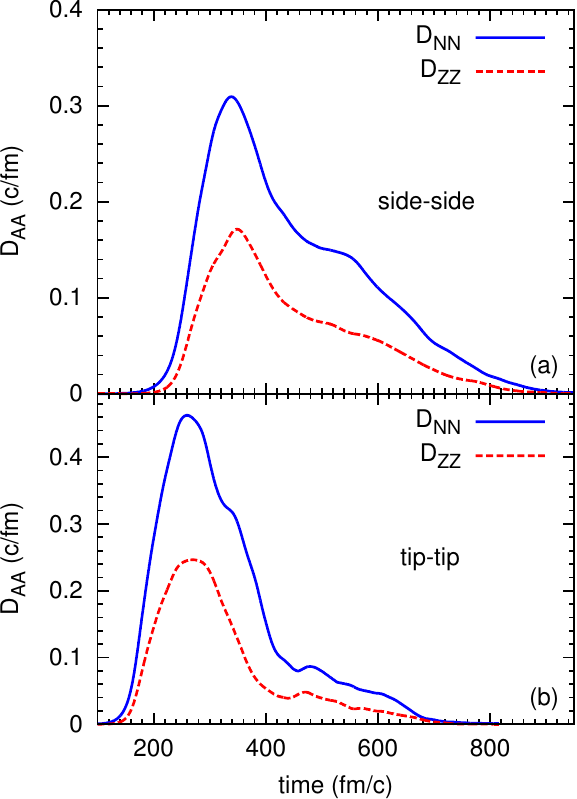}
\caption{(Color online) Neutron and proton diffusion coefficients in the central collisions of
$^{238}$U + $^{238}$U (a) side-side collision with energy $E_{c.m.} =900$~MeV, and (b) tip-tip
collision with energy $E_{c.m.} =1050$~MeV, respectively. \label{fig3}}
\end{figure}

In determining the primary fragment distributions, the main input quantities are the neutron and
proton diffusions coefficients given in Eq.~\eqref{eq18}. The diffusion coefficients are entirely
determined by the occupied time-dependent single-particle states. The TDHF theory includes the
one-body dissipation mechanism. We can use the same information provided by the TDHF to calculate
the diffusion coefficients which describe the fluctuation mechanism of the collective motion. 
The reason behind this fact is the fundamental relation that exists between dissipation and fluctuation 
mechanism of the collective motion as stated in the fluctuation-dissipation 
theorem~\cite{gardiner1991,weiss1999}. Fig.~\ref{fig3} shows the neutron (solid lines) and proton 
(dashed lines) diffusion coefficients in the
side-side and the tip-tip central collisions of $^{238}$U + $^{238}$U at bombarding energies
$E_{c.m.}=900$~MeV and  $E_{c.m.}=1050$~MeV, respectively. 
\begin{figure}[!hpt]
\includegraphics[width=7.7cm]{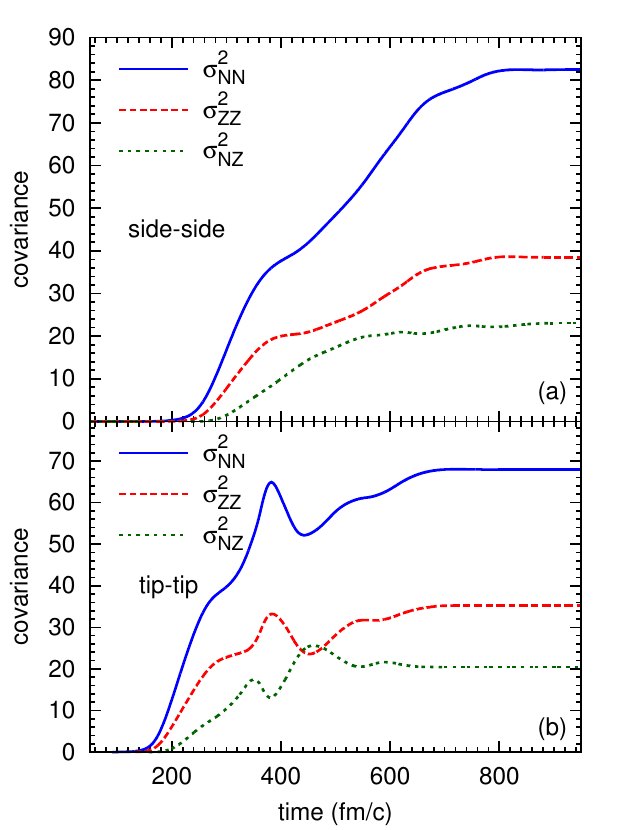}
\hspace{-0.7cm}
\caption{(Color online) Neutron, proton co-variances in the central collisions of $^{238}$U +
$^{238}$U (a) side-side collision with energy $E_{c.m.} =900$~MeV, and (b) tip-tip collision with
energy $E_{c.m.} =1050$~MeV, respectively.\label{fig4}}
\end{figure}
\begin{figure}[!hpt]
\includegraphics[width=7.5cm]{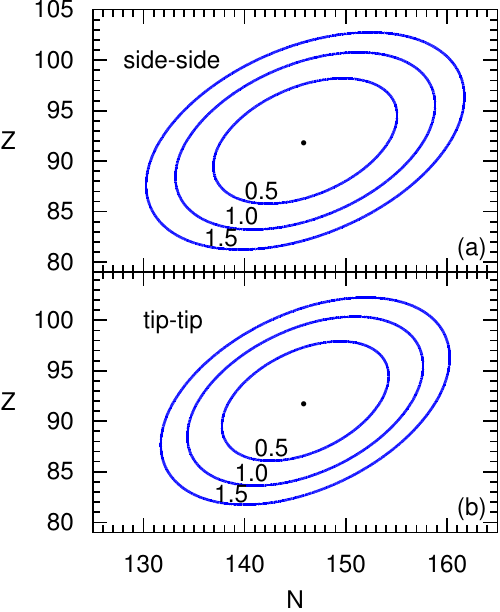}
\caption{(Color online) Equal probability lines for primary fragment formation with $C=0.5$, $1.0$,
$1.5$ in the central collisions of $^{238}$U + $^{238}$U (a) side-side collision with energy $E_{c.m.}
=900$~MeV, and (b) tip-tip collision with energy $E_{c.m.} =1050$~MeV, respectively. \label{fig5}}
\end{figure}
We determine
the proton, neutron co-variance by solving the coupled differential Eqs.~(\ref{eq9}-\ref{eq11}) with the initial
conditions $\sigma_{nn}(0)=0$, $\sigma_{pp}(0)=0$ and $\sigma_{np}(0)=0$. Fig.~\ref{fig4} illustrates these
co-variance as a function of time in the side-side and the tip-tip central collisions of $^{238}$U +
$^{238}$U. Primary fragment distribution in $N-Z$ plane is determined by a correlated Gaussian given
by Eq.~\eqref{eq12}. The elliptic curves in Fig.~\ref{fig5} show equal probability lines relative to the
center point for producing fragments for three values of the exponent $C=0.5$, $1.0$, $1.5$ in the
Gaussian function. For example the probability for producing fragments on the ellipse with $C=0.5$
relative to the symmetric fragmentation is $\exp(-0.5)=0.6$. Primary fragment distributions have a similar
behavior in both side-side and tip-tip collisions as seen from panels (a) and (b). The variance of
fragment mass distributions is determined by
\begin{align} \label{eq27}
\sigma_{AA}^{2}(t)=\sigma_{NN}^{2}(t)+\sigma_{ZZ}^{2}(t)+2\sigma _{NZ}^{2}(t).
\end{align}
As seen from Fig.~\ref{fig4}, at the end of the final states of collisions the co-variances of the fragment
mass distribution have the values $\sigma_{AA}(t)=12.9$ and $\sigma_{AA}(t)=12.0$ in side-side and
tip-tip collisions, respectively. Fig.~\ref{fig6} illustrates the Gaussian form of the mass distributions of
the primary fragments with a mean value $\overline{A}=238$ and variances $\sigma _{AA} (t)=12.9$ and
$\sigma _{AA}(t)=12.0$.

In the symmetric fragmentation of the final state, we can determine the excitation energy of each final
$^{238}$U nucleus by calculating the final total kinetic energy ($TKE$) from the TDHF solutions. 
We find $TKE=620$~MeV and $TKE=634$~MeV in the side-side and the tip-tip collisions, respectively.
From the energy conservation, $E^*=E_{cm} -TKE$, we find that the excitation energy of each $^{238}$U
nucleus is $E^*=140$~MeV and $E^*=208$~MeV, in the side-side and the tip-tip collisions.
As a result of multi-nucleon transfer in the collisions, there are many binary fragments in the
final state as indicated in distributions in Fig.~\ref{fig5}.
\begin{figure}[!hpt]
\includegraphics[width=8cm]{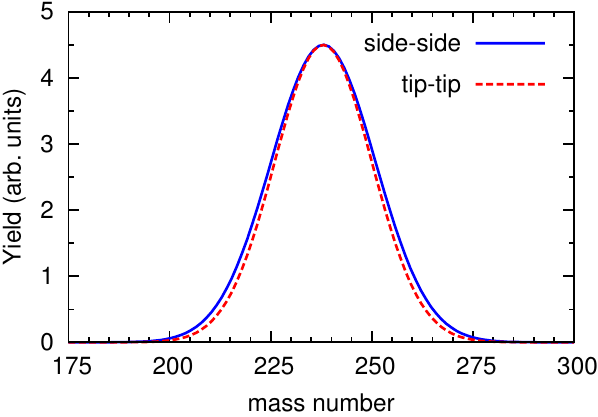}
\caption{(Color online) Primary fragment mass distributions in the central collisions of $^{238}$U +
$^{238}$U at side-side collision with energy $E_{c.m.} =900$~MeV (solid line) and at tip-tip
collision with energy $E_{c.m.} =1050$~MeV (dashed line). \label{fig6}}
\end{figure}
In the present work, we cannot calculate the excitation energies of each final fragment pair, but we
can estimate them by using the Viola systematics. It is very reasonable to assume that all available initial
relative kinetic energy is dissipated into the internal excitations and is shared between the
fragments in proportion to the ratio of masses in possible final binary channel. According to the
Viola formula, total excitation $E_{c}^{*}$ in a binary channel is determined by $E_{c}^{*} =E_{cm}
+Q_{c} -(TKE)_{c}$. Here, $Q_{c}$ is the Q-value of the binary channel and $(TKE)_{c}$ indicates the
total final kinetic energy of fragments. $(TKE)_{c}$ is approximately determined by Coulomb
potential energy of the binary fragments at an effective relative distance determined by an
adjustable parameter $r_{0}$ as,
\begin{align} \label{eq28}
(TKE)_{c} =\frac{1}{4\pi\varepsilon_{0}} \frac{Z_{1c} Z_{2c}e^2}{r_{0} \left(A_{1c}^{1/3} +A_{2c}^{1/3} \right)}.
\end{align}
With help of $TKE$ of the symmetric binary channel, we adjust the parameter $r_{0} =1.59$ fm and
$r_{0} =1.55$ fm for the side-side and tip-tip collisions, respectively. We estimate that primary
fragments inside the eliptic region with $C=1.5$ have excitation energies in the range of $(120-150)$~MeV 
and $(185-225)$~MeV in the
side-side and tip-tip collisions, respectively.  Highly exited intermediate mass fragments cool
down by particle evaporations and heavy-fragments should immediately fission. However we do not
perform de-excitation calculations of the primary fragments in this work.

\section{Conclusions}
The SMF approach improves the standard mean-field description by incorporating thermal and quantal
fluctuations in the collective motion. The approach requires to generate an ensemble of mean field
trajectories. The initial conditions for the events in the ensemble are specified by the quantal and
thermal fluctuations in the initial state in a suitable manner, and each event is evolved by its own
self-consistent mean-field Hamiltonian. In reactions where the colliding system maintains a
di-nuclear structure, the reaction dynamics can be described in terms of a set of relevant
macroscopic variables, which can be defined with the help of the window dynamics. The SMF approach
gives rise to a quantal Langevin description for the evolution of the macroscopic variables. In this
work, we apply this approach and analyze multi-nucleon transfer mechanism in the central collisions of
$^{238}$U + $^{238}$U in side-side geometry with energy $E_{c.m.} =900$~MeV and in tip-tip geometry
with energy $E_{c.m.} =1050$~MeV. Fluctuation mechanism of neutron and proton exchanges is
described by the quantal diffusion coefficients. Quantal diffusion coefficients are entirely
determined by the single-particle states of the TDHF equations. These coefficients include the full
geometry of the collision process and the effect of the shell structure. They do not involve any
adjustable parameters and do not require any additional information. Deep underlying reason behind
this is the fact that the dissipation and fluctuation aspects of the dynamics are connected according to
the fluctuation-dissipation theorem of non-equilibrium statistical mechanics. We estimate the
excitation energies of the primary fragments with the help of Viola formula which provides an
approximate description of the total final kinetic energy of the binary fragments. The
highly excited fragments are cooled down by particle emission and in particular highly excited heavy
fragments are expected to decay rapidly by fission. We plan to carry out de-excitation calculations
and determine the secondary fragment distributions in a subsequent work.

\begin{acknowledgments}
S.A. gratefully acknowledges the IPN-Orsay and the Middle East Technical University for warm
hospitality extended to him during his visits. S.A. also gratefully acknowledges useful discussions
with D. Lacroix, and very much thankful to F. Ayik for continuous support and encouragement. This
work is supported in part by US DOE Grant Nos. DE-SC0015513 and DE-SC0013847.
\end{acknowledgments}

\bibliography{VU_bibtex_master}

\begin{thebibliography}{56}%
\makeatletter
\providecommand \@ifxundefined [1]{%
 \@ifx{#1\undefined}
}%
\providecommand \@ifnum [1]{%
 \ifnum #1\expandafter \@firstoftwo
 \else \expandafter \@secondoftwo
 \fi
}%
\providecommand \@ifx [1]{%
 \ifx #1\expandafter \@firstoftwo
 \else \expandafter \@secondoftwo
 \fi
}%
\providecommand \natexlab [1]{#1}%
\providecommand \enquote  [1]{``#1''}%
\providecommand \bibnamefont  [1]{#1}%
\providecommand \bibfnamefont [1]{#1}%
\providecommand \citenamefont [1]{#1}%
\providecommand \href@noop [0]{\@secondoftwo}%
\providecommand \href [0]{\begingroup \@sanitize@url \@href}%
\providecommand \@href[1]{\@@startlink{#1}\@@href}%
\providecommand \@@href[1]{\endgroup#1\@@endlink}%
\providecommand \@sanitize@url [0]{\catcode `\\12\catcode `\$12\catcode
  `\&12\catcode `\#12\catcode `\^12\catcode `\_12\catcode `\%12\relax}%
\providecommand \@@startlink[1]{}%
\providecommand \@@endlink[0]{}%
\providecommand \url  [0]{\begingroup\@sanitize@url \@url }%
\providecommand \@url [1]{\endgroup\@href {#1}{\urlprefix }}%
\providecommand \urlprefix  [0]{URL }%
\providecommand \Eprint [0]{\href }%
\providecommand \doibase [0]{http://dx.doi.org/}%
\providecommand \selectlanguage [0]{\@gobble}%
\providecommand \bibinfo  [0]{\@secondoftwo}%
\providecommand \bibfield  [0]{\@secondoftwo}%
\providecommand \translation [1]{[#1]}%
\providecommand \BibitemOpen [0]{}%
\providecommand \bibitemStop [0]{}%
\providecommand \bibitemNoStop [0]{.\EOS\space}%
\providecommand \EOS [0]{\spacefactor3000\relax}%
\providecommand \BibitemShut  [1]{\csname bibitem#1\endcsname}%
\let\auto@bib@innerbib\@empty
\bibitem [{\citenamefont {Adamian}\ \emph {et~al.}(2003)\citenamefont
  {Adamian}, \citenamefont {Antonenko},\ and\ \citenamefont
  {Scheid}}]{adamian2003}%
  \BibitemOpen
  \bibfield  {author} {\bibinfo {author} {\bibfnamefont {G.~G.}\ \bibnamefont
  {Adamian}}, \bibinfo {author} {\bibfnamefont {N.~V.}\ \bibnamefont
  {Antonenko}}, \ and\ \bibinfo {author} {\bibfnamefont {W.}~\bibnamefont
  {Scheid}},\ }\bibfield  {title} {\enquote {\bibinfo {title}
  {{C}haracteristics of quasifission products within the dinuclear system
  model},}\ }\href {\doibase 10.1103/PhysRevC.68.034601} {\bibfield  {journal}
  {\bibinfo  {journal} {Phys. Rev. C}\ }\textbf {\bibinfo {volume} {68}},\
  \bibinfo {pages} {034601} (\bibinfo {year} {2003})}\BibitemShut {NoStop}%
\bibitem [{\citenamefont {{Valery Zagrebaev}}\ and\ \citenamefont {{Walter
  Greiner}}(2007)}]{zagrebaev2007}%
  \BibitemOpen
  \bibfield  {author} {\bibinfo {author} {\bibnamefont {{Valery Zagrebaev}}}\
  and\ \bibinfo {author} {\bibnamefont {{Walter Greiner}}},\ }\bibfield
  {title} {\enquote {\bibinfo {title} {{S}hell effects in damped collisions: a
  new way to superheavies},}\ }\href {\doibase 10.1088/0954-3899/34/11/004}
  {\bibfield  {journal} {\bibinfo  {journal} {J. Phys. G}\ }\textbf {\bibinfo
  {volume} {34}},\ \bibinfo {pages} {2265} (\bibinfo {year}
  {2007})}\BibitemShut {NoStop}%
\bibitem [{\citenamefont {Aritomo}(2009)}]{aritomo2009}%
  \BibitemOpen
  \bibfield  {author} {\bibinfo {author} {\bibfnamefont {Y.}~\bibnamefont
  {Aritomo}},\ }\bibfield  {title} {\enquote {\bibinfo {title} {{A}nalysis of
  dynamical processes using the mass distribution of fission fragments in
  heavy-ion reactions},}\ }\href {\doibase 10.1103/PhysRevC.80.064604}
  {\bibfield  {journal} {\bibinfo  {journal} {Phys. Rev. C}\ }\textbf {\bibinfo
  {volume} {80}},\ \bibinfo {pages} {064604} (\bibinfo {year}
  {2009})}\BibitemShut {NoStop}%
\bibitem [{\citenamefont {Zhao}\ \emph {et~al.}(2016)\citenamefont {Zhao},
  \citenamefont {Li}, \citenamefont {Zhang}, \citenamefont {Wang},
  \citenamefont {Li}, \citenamefont {Shen}, \citenamefont {Wang},\ and\
  \citenamefont {Wu}}]{zhao2016}%
  \BibitemOpen
  \bibfield  {author} {\bibinfo {author} {\bibfnamefont {Kai}\ \bibnamefont
  {Zhao}}, \bibinfo {author} {\bibfnamefont {Zhuxia}\ \bibnamefont {Li}},
  \bibinfo {author} {\bibfnamefont {Yingxun}\ \bibnamefont {Zhang}}, \bibinfo
  {author} {\bibfnamefont {Ning}\ \bibnamefont {Wang}}, \bibinfo {author}
  {\bibfnamefont {Qingfeng}\ \bibnamefont {Li}}, \bibinfo {author}
  {\bibfnamefont {Caiwan}\ \bibnamefont {Shen}}, \bibinfo {author}
  {\bibfnamefont {Yongjia}\ \bibnamefont {Wang}}, \ and\ \bibinfo {author}
  {\bibfnamefont {Xizhen}\ \bibnamefont {Wu}},\ }\bibfield  {title} {\enquote
  {\bibinfo {title} {Production of unknown neutron--rich isotopes in
  $^{238}${U}+$^{238}${U} collisions at near--barrier energy},}\ }\href
  {\doibase 10.1103/PhysRevC.94.024601} {\bibfield  {journal} {\bibinfo
  {journal} {Phys. Rev. C}\ }\textbf {\bibinfo {volume} {94}},\ \bibinfo
  {pages} {024601} (\bibinfo {year} {2016})}\BibitemShut {NoStop}%
\bibitem [{\citenamefont {Negele}(1982)}]{negele1982}%
  \BibitemOpen
  \bibfield  {author} {\bibinfo {author} {\bibfnamefont {J.~W.}\ \bibnamefont
  {Negele}},\ }\bibfield  {title} {\enquote {\bibinfo {title} {{T}he mean-field
  theory of nuclear-structure and dynamics},}\ }\href {\doibase
  10.1103/RevModPhys.54.913} {\bibfield  {journal} {\bibinfo  {journal} {Rev.
  Mod. Phys.}\ }\textbf {\bibinfo {volume} {54}},\ \bibinfo {pages} {913--1015}
  (\bibinfo {year} {1982})}\BibitemShut {NoStop}%
\bibitem [{\citenamefont {Nakatsukasa}\ \emph {et~al.}(2016)\citenamefont
  {Nakatsukasa}, \citenamefont {Matsuyanagi}, \citenamefont {Matsuo},\ and\
  \citenamefont {Yabana}}]{nakatsukasa2016}%
  \BibitemOpen
  \bibfield  {author} {\bibinfo {author} {\bibfnamefont {Takashi}\ \bibnamefont
  {Nakatsukasa}}, \bibinfo {author} {\bibfnamefont {Kenichi}\ \bibnamefont
  {Matsuyanagi}}, \bibinfo {author} {\bibfnamefont {Masayuki}\ \bibnamefont
  {Matsuo}}, \ and\ \bibinfo {author} {\bibfnamefont {Kazuhiro}\ \bibnamefont
  {Yabana}},\ }\bibfield  {title} {\enquote {\bibinfo {title} {Time-dependent
  density-functional description of nuclear dynamics},}\ }\href {\doibase
  10.1103/RevModPhys.88.045004} {\bibfield  {journal} {\bibinfo  {journal}
  {Rev. Mod. Phys.}\ }\textbf {\bibinfo {volume} {88}},\ \bibinfo {pages}
  {045004} (\bibinfo {year} {2016})}\BibitemShut {NoStop}%
\bibitem [{\citenamefont {Simenel}(2012)}]{simenel2012}%
  \BibitemOpen
  \bibfield  {author} {\bibinfo {author} {\bibfnamefont {C\'edric}\
  \bibnamefont {Simenel}},\ }\bibfield  {title} {\enquote {\bibinfo {title}
  {{N}uclear quantum many-body dynamics},}\ }\href {\doibase
  10.1140/epja/i2012-12152-0} {\bibfield  {journal} {\bibinfo  {journal} {Eur.
  Phys. J. A}\ }\textbf {\bibinfo {volume} {48}},\ \bibinfo {pages} {152}
  (\bibinfo {year} {2012})}\BibitemShut {NoStop}%
\bibitem [{\citenamefont {{C\'edric Golabek}}\ and\ \citenamefont {{C\'edric
  Simenel}}(2009)}]{golabek2009}%
  \BibitemOpen
  \bibfield  {author} {\bibinfo {author} {\bibnamefont {{C\'edric Golabek}}}\
  and\ \bibinfo {author} {\bibnamefont {{C\'edric Simenel}}},\ }\bibfield
  {title} {\enquote {\bibinfo {title} {{C}ollision {D}ynamics of {T}wo
  $^{238}${U A}tomic {N}uclei},}\ }\href {\doibase
  10.1103/PhysRevLett.103.042701} {\bibfield  {journal} {\bibinfo  {journal}
  {Phys. Rev. Lett.}\ }\textbf {\bibinfo {volume} {103}},\ \bibinfo {pages}
  {042701} (\bibinfo {year} {2009})}\BibitemShut {NoStop}%
\bibitem [{\citenamefont {{David J. Kedziora}}\ and\ \citenamefont {{C\'edric
  Simenel}}(2010)}]{kedziora2010}%
  \BibitemOpen
  \bibfield  {author} {\bibinfo {author} {\bibnamefont {{David J. Kedziora}}}\
  and\ \bibinfo {author} {\bibnamefont {{C\'edric Simenel}}},\ }\bibfield
  {title} {\enquote {\bibinfo {title} {{N}ew inverse quasifission mechanism to
  produce neutron-rich transfermium nuclei},}\ }\href {\doibase
  10.1103/PhysRevC.81.044613} {\bibfield  {journal} {\bibinfo  {journal} {Phys.
  Rev. C}\ }\textbf {\bibinfo {volume} {81}},\ \bibinfo {pages} {044613}
  (\bibinfo {year} {2010})}\BibitemShut {NoStop}%
\bibitem [{\citenamefont {Simenel}\ \emph {et~al.}(2012)\citenamefont
  {Simenel}, \citenamefont {Hinde}, \citenamefont {{du Rietz}}, \citenamefont
  {Dasgupta}, \citenamefont {Evers}, \citenamefont {Lin}, \citenamefont
  {Luong},\ and\ \citenamefont {Wakhle}}]{simenel2012b}%
  \BibitemOpen
  \bibfield  {author} {\bibinfo {author} {\bibfnamefont {C.}~\bibnamefont
  {Simenel}}, \bibinfo {author} {\bibfnamefont {D.~J.}\ \bibnamefont {Hinde}},
  \bibinfo {author} {\bibfnamefont {R.}~\bibnamefont {{du Rietz}}}, \bibinfo
  {author} {\bibfnamefont {M.}~\bibnamefont {Dasgupta}}, \bibinfo {author}
  {\bibfnamefont {M.}~\bibnamefont {Evers}}, \bibinfo {author} {\bibfnamefont
  {C.~J.}\ \bibnamefont {Lin}}, \bibinfo {author} {\bibfnamefont {D.~H.}\
  \bibnamefont {Luong}}, \ and\ \bibinfo {author} {\bibfnamefont
  {A.}~\bibnamefont {Wakhle}},\ }\bibfield  {title} {\enquote {\bibinfo {title}
  {{I}nfluence of entrance-channel magicity and isospin on quasi-fission},}\
  }\href {\doibase 10.1016/j.physletb.2012.03.063} {\bibfield  {journal}
  {\bibinfo  {journal} {Phys. Lett. B}\ }\textbf {\bibinfo {volume} {710}},\
  \bibinfo {pages} {607--611} (\bibinfo {year} {2012})}\BibitemShut {NoStop}%
\bibitem [{\citenamefont {Wakhle}\ \emph {et~al.}(2014)\citenamefont {Wakhle},
  \citenamefont {Simenel}, \citenamefont {Hinde}, \citenamefont {Dasgupta},
  \citenamefont {Evers}, \citenamefont {Luong}, \citenamefont {du~Rietz},\ and\
  \citenamefont {Williams}}]{wakhle2014}%
  \BibitemOpen
  \bibfield  {author} {\bibinfo {author} {\bibfnamefont {A.}~\bibnamefont
  {Wakhle}}, \bibinfo {author} {\bibfnamefont {C.}~\bibnamefont {Simenel}},
  \bibinfo {author} {\bibfnamefont {D.~J.}\ \bibnamefont {Hinde}}, \bibinfo
  {author} {\bibfnamefont {M.}~\bibnamefont {Dasgupta}}, \bibinfo {author}
  {\bibfnamefont {M.}~\bibnamefont {Evers}}, \bibinfo {author} {\bibfnamefont
  {D.~H.}\ \bibnamefont {Luong}}, \bibinfo {author} {\bibfnamefont
  {R.}~\bibnamefont {du~Rietz}}, \ and\ \bibinfo {author} {\bibfnamefont
  {E.}~\bibnamefont {Williams}},\ }\bibfield  {title} {\enquote {\bibinfo
  {title} {{I}nterplay between {Q}uantum {S}hells and {O}rientation in
  {Q}uasifission},}\ }\href {\doibase 10.1103/PhysRevLett.113.182502}
  {\bibfield  {journal} {\bibinfo  {journal} {Phys. Rev. Lett.}\ }\textbf
  {\bibinfo {volume} {113}},\ \bibinfo {pages} {182502} (\bibinfo {year}
  {2014})}\BibitemShut {NoStop}%
\bibitem [{\citenamefont {Oberacker}\ \emph {et~al.}(2014)\citenamefont
  {Oberacker}, \citenamefont {Umar},\ and\ \citenamefont
  {Simenel}}]{oberacker2014}%
  \BibitemOpen
  \bibfield  {author} {\bibinfo {author} {\bibfnamefont {V.~E.}\ \bibnamefont
  {Oberacker}}, \bibinfo {author} {\bibfnamefont {A.~S.}\ \bibnamefont {Umar}},
  \ and\ \bibinfo {author} {\bibfnamefont {C.}~\bibnamefont {Simenel}},\
  }\bibfield  {title} {\enquote {\bibinfo {title} {{D}issipative dynamics in
  quasifission},}\ }\href {\doibase 10.1103/PhysRevC.90.054605} {\bibfield
  {journal} {\bibinfo  {journal} {Phys. Rev. C}\ }\textbf {\bibinfo {volume}
  {90}},\ \bibinfo {pages} {054605} (\bibinfo {year} {2014})}\BibitemShut
  {NoStop}%
\bibitem [{\citenamefont {Hammerton}\ \emph {et~al.}(2015)\citenamefont
  {Hammerton}, \citenamefont {Kohley}, \citenamefont {Hinde}, \citenamefont
  {Dasgupta}, \citenamefont {Wakhle}, \citenamefont {Williams}, \citenamefont
  {Oberacker}, \citenamefont {Umar}, \citenamefont {Carter}, \citenamefont
  {Cook}, \citenamefont {Greene}, \citenamefont {Jeung}, \citenamefont {Luong},
  \citenamefont {McNeil}, \citenamefont {Palshetkar}, \citenamefont {Rafferty},
  \citenamefont {Simenel},\ and\ \citenamefont {Stiefel}}]{hammerton2015}%
  \BibitemOpen
  \bibfield  {author} {\bibinfo {author} {\bibfnamefont {K.}~\bibnamefont
  {Hammerton}}, \bibinfo {author} {\bibfnamefont {Z.}~\bibnamefont {Kohley}},
  \bibinfo {author} {\bibfnamefont {D.~J.}\ \bibnamefont {Hinde}}, \bibinfo
  {author} {\bibfnamefont {M.}~\bibnamefont {Dasgupta}}, \bibinfo {author}
  {\bibfnamefont {A.}~\bibnamefont {Wakhle}}, \bibinfo {author} {\bibfnamefont
  {E.}~\bibnamefont {Williams}}, \bibinfo {author} {\bibfnamefont {V.~E.}\
  \bibnamefont {Oberacker}}, \bibinfo {author} {\bibfnamefont {A.~S.}\
  \bibnamefont {Umar}}, \bibinfo {author} {\bibfnamefont {I.~P.}\ \bibnamefont
  {Carter}}, \bibinfo {author} {\bibfnamefont {K.~J.}\ \bibnamefont {Cook}},
  \bibinfo {author} {\bibfnamefont {J.}~\bibnamefont {Greene}}, \bibinfo
  {author} {\bibfnamefont {D.~Y.}\ \bibnamefont {Jeung}}, \bibinfo {author}
  {\bibfnamefont {D.~H.}\ \bibnamefont {Luong}}, \bibinfo {author}
  {\bibfnamefont {S.~D.}\ \bibnamefont {McNeil}}, \bibinfo {author}
  {\bibfnamefont {C.~S.}\ \bibnamefont {Palshetkar}}, \bibinfo {author}
  {\bibfnamefont {D.~C.}\ \bibnamefont {Rafferty}}, \bibinfo {author}
  {\bibfnamefont {C.}~\bibnamefont {Simenel}}, \ and\ \bibinfo {author}
  {\bibfnamefont {K.}~\bibnamefont {Stiefel}},\ }\bibfield  {title} {\enquote
  {\bibinfo {title} {{R}educed quasifission competition in fusion reactions
  forming neutron-rich heavy elements},}\ }\href {\doibase
  10.1103/PhysRevC.91.041602} {\bibfield  {journal} {\bibinfo  {journal} {Phys.
  Rev. C}\ }\textbf {\bibinfo {volume} {91}},\ \bibinfo {pages} {041602(R)}
  (\bibinfo {year} {2015})}\BibitemShut {NoStop}%
\bibitem [{\citenamefont {Umar}\ and\ \citenamefont
  {Oberacker}(2015)}]{umar2015c}%
  \BibitemOpen
  \bibfield  {author} {\bibinfo {author} {\bibfnamefont {A.~S.}\ \bibnamefont
  {Umar}}\ and\ \bibinfo {author} {\bibfnamefont {V.~E.}\ \bibnamefont
  {Oberacker}},\ }\bibfield  {title} {\enquote {\bibinfo {title}
  {{T}ime-dependent {HF} approach to {SHE} dynamics},}\ }\href {\doibase
  10.1016/j.nuclphysa.2015.02.011} {\bibfield  {journal} {\bibinfo  {journal}
  {Nucl. Phys. A}\ }\textbf {\bibinfo {volume} {944}},\ \bibinfo {pages}
  {238--256} (\bibinfo {year} {2015})}\BibitemShut {NoStop}%
\bibitem [{\citenamefont {Umar}\ \emph {et~al.}(2015)\citenamefont {Umar},
  \citenamefont {Oberacker},\ and\ \citenamefont {Simenel}}]{umar2015a}%
  \BibitemOpen
  \bibfield  {author} {\bibinfo {author} {\bibfnamefont {A.~S.}\ \bibnamefont
  {Umar}}, \bibinfo {author} {\bibfnamefont {V.~E.}\ \bibnamefont {Oberacker}},
  \ and\ \bibinfo {author} {\bibfnamefont {C.}~\bibnamefont {Simenel}},\
  }\bibfield  {title} {\enquote {\bibinfo {title} {{S}hape evolution and
  collective dynamics of quasifission in the time-dependent {H}artree-{F}ock
  approach},}\ }\href {\doibase 10.1103/PhysRevC.92.024621} {\bibfield
  {journal} {\bibinfo  {journal} {Phys. Rev. C}\ }\textbf {\bibinfo {volume}
  {92}},\ \bibinfo {pages} {024621} (\bibinfo {year} {2015})}\BibitemShut
  {NoStop}%
\bibitem [{\citenamefont {Sekizawa}\ and\ \citenamefont
  {Yabana}(2016)}]{sekizawa2016}%
  \BibitemOpen
  \bibfield  {author} {\bibinfo {author} {\bibfnamefont {Kazuyuki}\
  \bibnamefont {Sekizawa}}\ and\ \bibinfo {author} {\bibfnamefont {Kazuhiro}\
  \bibnamefont {Yabana}},\ }\bibfield  {title} {\enquote {\bibinfo {title}
  {{T}ime-dependent {H}artree-{F}ock calculations for multinucleon transfer and
  quasifission processes in the $^{64}\text{Ni}+^{238}\text{U}$ reaction},}\
  }\href {\doibase 10.1103/PhysRevC.93.054616} {\bibfield  {journal} {\bibinfo
  {journal} {Phys. Rev. C}\ }\textbf {\bibinfo {volume} {93}},\ \bibinfo
  {pages} {054616} (\bibinfo {year} {2016})}\BibitemShut {NoStop}%
\bibitem [{\citenamefont {Umar}\ \emph {et~al.}(2016)\citenamefont {Umar},
  \citenamefont {Oberacker},\ and\ \citenamefont {Simenel}}]{umar2016}%
  \BibitemOpen
  \bibfield  {author} {\bibinfo {author} {\bibfnamefont {A.~S.}\ \bibnamefont
  {Umar}}, \bibinfo {author} {\bibfnamefont {V.~E.}\ \bibnamefont {Oberacker}},
  \ and\ \bibinfo {author} {\bibfnamefont {C.}~\bibnamefont {Simenel}},\
  }\bibfield  {title} {\enquote {\bibinfo {title} {Fusion and quasifission
  dynamics in the reactions $^{48}\mathrm{Ca}+^{249}\mathrm{Bk}$ and
  $^{50}\mathrm{Ti}+^{249}\mathrm{Bk}$ using a time-dependent {H}artree-{F}ock
  approach},}\ }\href {\doibase 10.1103/PhysRevC.94.024605} {\bibfield
  {journal} {\bibinfo  {journal} {Phys. Rev. C}\ }\textbf {\bibinfo {volume}
  {94}},\ \bibinfo {pages} {024605} (\bibinfo {year} {2016})}\BibitemShut
  {NoStop}%
\bibitem [{\citenamefont {Simenel}\ and\ \citenamefont
  {Umar}(2014)}]{simenel2014a}%
  \BibitemOpen
  \bibfield  {author} {\bibinfo {author} {\bibfnamefont {C.}~\bibnamefont
  {Simenel}}\ and\ \bibinfo {author} {\bibfnamefont {A.~S.}\ \bibnamefont
  {Umar}},\ }\bibfield  {title} {\enquote {\bibinfo {title} {{F}ormation and
  dynamics of fission fragments},}\ }\href {\doibase
  10.1103/PhysRevC.89.031601} {\bibfield  {journal} {\bibinfo  {journal} {Phys.
  Rev. C}\ }\textbf {\bibinfo {volume} {89}},\ \bibinfo {pages} {031601(R)}
  (\bibinfo {year} {2014})}\BibitemShut {NoStop}%
\bibitem [{\citenamefont {Scamps}\ \emph {et~al.}(2015)\citenamefont {Scamps},
  \citenamefont {Simenel},\ and\ \citenamefont {Lacroix}}]{scamps2015a}%
  \BibitemOpen
  \bibfield  {author} {\bibinfo {author} {\bibfnamefont {Guillaume}\
  \bibnamefont {Scamps}}, \bibinfo {author} {\bibfnamefont {C\'edric}\
  \bibnamefont {Simenel}}, \ and\ \bibinfo {author} {\bibfnamefont {Denis}\
  \bibnamefont {Lacroix}},\ }\bibfield  {title} {\enquote {\bibinfo {title}
  {{S}uperfluid dynamics of $^{258}\mathrm{Fm}$ fission},}\ }\href {\doibase
  10.1103/PhysRevC.92.011602} {\bibfield  {journal} {\bibinfo  {journal} {Phys.
  Rev. C}\ }\textbf {\bibinfo {volume} {92}},\ \bibinfo {pages} {011602}
  (\bibinfo {year} {2015})}\BibitemShut {NoStop}%
\bibitem [{\citenamefont {Simenel}\ \emph {et~al.}(2016)\citenamefont
  {Simenel}, \citenamefont {Scamps}, \citenamefont {Lacroix},\ and\
  \citenamefont {Umar}}]{simenel2016a}%
  \BibitemOpen
  \bibfield  {author} {\bibinfo {author} {\bibfnamefont {C.}~\bibnamefont
  {Simenel}}, \bibinfo {author} {\bibfnamefont {G.}~\bibnamefont {Scamps}},
  \bibinfo {author} {\bibfnamefont {D.}~\bibnamefont {Lacroix}}, \ and\
  \bibinfo {author} {\bibfnamefont {A.~S.}\ \bibnamefont {Umar}},\ }\bibfield
  {title} {\enquote {\bibinfo {title} {Superfluid fission dynamics with
  microscopic approaches},}\ }\href {\doibase 10.1051/epjconf/201610707001}
  {\bibfield  {journal} {\bibinfo  {journal} {{EPJ} Web. Conf.}\ }\textbf
  {\bibinfo {volume} {107}},\ \bibinfo {pages} {07001} (\bibinfo {year}
  {2016})}\BibitemShut {NoStop}%
\bibitem [{\citenamefont {Goddard}\ \emph {et~al.}(2015)\citenamefont
  {Goddard}, \citenamefont {Stevenson},\ and\ \citenamefont
  {Rios}}]{goddard2015}%
  \BibitemOpen
  \bibfield  {author} {\bibinfo {author} {\bibfnamefont {P.~M.}\ \bibnamefont
  {Goddard}}, \bibinfo {author} {\bibfnamefont {P.~D.}\ \bibnamefont
  {Stevenson}}, \ and\ \bibinfo {author} {\bibfnamefont {A.}~\bibnamefont
  {Rios}},\ }\bibfield  {title} {\enquote {\bibinfo {title} {{F}ission dynamics
  within time-dependent {H}artree-{F}ock: deformation-induced fission},}\
  }\href {\doibase 10.1103/PhysRevC.92.054610} {\bibfield  {journal} {\bibinfo
  {journal} {Phys. Rev. C}\ }\textbf {\bibinfo {volume} {92}},\ \bibinfo
  {pages} {054610} (\bibinfo {year} {2015})}\BibitemShut {NoStop}%
\bibitem [{\citenamefont {Goddard}\ \emph {et~al.}(2016)\citenamefont
  {Goddard}, \citenamefont {Stevenson},\ and\ \citenamefont
  {Rios}}]{goddard2016}%
  \BibitemOpen
  \bibfield  {author} {\bibinfo {author} {\bibfnamefont {P.~M.}\ \bibnamefont
  {Goddard}}, \bibinfo {author} {\bibfnamefont {P.~D.}\ \bibnamefont
  {Stevenson}}, \ and\ \bibinfo {author} {\bibfnamefont {A.}~\bibnamefont
  {Rios}},\ }\bibfield  {title} {\enquote {\bibinfo {title} {Fission dynamics
  within time--dependent {H}artree--{F}ock. {II}. {B}oost-induced fission},}\
  }\href {\doibase 10.1103/PhysRevC.93.014620} {\bibfield  {journal} {\bibinfo
  {journal} {Phys. Rev. C}\ }\textbf {\bibinfo {volume} {93}},\ \bibinfo
  {pages} {014620} (\bibinfo {year} {2016})}\BibitemShut {NoStop}%
\bibitem [{\citenamefont {Bulgac}\ \emph {et~al.}(2016)\citenamefont {Bulgac},
  \citenamefont {Magierski}, \citenamefont {Roche},\ and\ \citenamefont
  {Stetcu}}]{bulgac2016}%
  \BibitemOpen
  \bibfield  {author} {\bibinfo {author} {\bibfnamefont {Aurel}\ \bibnamefont
  {Bulgac}}, \bibinfo {author} {\bibfnamefont {Piotr}\ \bibnamefont
  {Magierski}}, \bibinfo {author} {\bibfnamefont {Kenneth~J.}\ \bibnamefont
  {Roche}}, \ and\ \bibinfo {author} {\bibfnamefont {Ionel}\ \bibnamefont
  {Stetcu}},\ }\bibfield  {title} {\enquote {\bibinfo {title} {{I}nduced
  {F}ission of $^{240}${P}u within a {R}eal-{T}ime {M}icroscopic
  {F}ramework},}\ }\href {\doibase 10.1103/physrevlett.116.122504} {\bibfield
  {journal} {\bibinfo  {journal} {Phys. Rev. Lett.}\ }\textbf {\bibinfo
  {volume} {116}},\ \bibinfo {pages} {122504} (\bibinfo {year}
  {2016})}\BibitemShut {NoStop}%
\bibitem [{\citenamefont {Umar}\ and\ \citenamefont
  {Oberacker}(2006)}]{umar2006c}%
  \BibitemOpen
  \bibfield  {author} {\bibinfo {author} {\bibfnamefont {A.~S.}\ \bibnamefont
  {Umar}}\ and\ \bibinfo {author} {\bibfnamefont {V.~E.}\ \bibnamefont
  {Oberacker}},\ }\bibfield  {title} {\enquote {\bibinfo {title}
  {{T}hree-dimensional unrestricted time-dependent {H}artree-{F}ock fusion
  calculations using the full {S}kyrme interaction},}\ }\href {\doibase
  10.1103/PhysRevC.73.054607} {\bibfield  {journal} {\bibinfo  {journal} {Phys.
  Rev. C}\ }\textbf {\bibinfo {volume} {73}},\ \bibinfo {pages} {054607}
  (\bibinfo {year} {2006})}\BibitemShut {NoStop}%
\bibitem [{\citenamefont {Maruhn}\ \emph {et~al.}(2014)\citenamefont {Maruhn},
  \citenamefont {Reinhard}, \citenamefont {Stevenson},\ and\ \citenamefont
  {Umar}}]{maruhn2014}%
  \BibitemOpen
  \bibfield  {author} {\bibinfo {author} {\bibfnamefont {J.~A.}\ \bibnamefont
  {Maruhn}}, \bibinfo {author} {\bibfnamefont {P.-G.}\ \bibnamefont
  {Reinhard}}, \bibinfo {author} {\bibfnamefont {P.~D.}\ \bibnamefont
  {Stevenson}}, \ and\ \bibinfo {author} {\bibfnamefont {A.~S.}\ \bibnamefont
  {Umar}},\ }\bibfield  {title} {\enquote {\bibinfo {title} {{T}he {TDHF C}ode
  {S}ky3{D}},}\ }\href {\doibase 10.1016/j.cpc.2014.04.008} {\bibfield
  {journal} {\bibinfo  {journal} {Comp. Phys. Comm.}\ }\textbf {\bibinfo
  {volume} {185}},\ \bibinfo {pages} {2195--2216} (\bibinfo {year}
  {2014})}\BibitemShut {NoStop}%
\bibitem [{\citenamefont {Schuetrumpf}\ \emph {et~al.}(2016)\citenamefont
  {Schuetrumpf}, \citenamefont {Nazarewicz},\ and\ \citenamefont
  {Reinhard}}]{schuetrumpf2016}%
  \BibitemOpen
  \bibfield  {author} {\bibinfo {author} {\bibfnamefont {B.}~\bibnamefont
  {Schuetrumpf}}, \bibinfo {author} {\bibfnamefont {W.}~\bibnamefont
  {Nazarewicz}}, \ and\ \bibinfo {author} {\bibfnamefont {P.-G.}\ \bibnamefont
  {Reinhard}},\ }\bibfield  {title} {\enquote {\bibinfo {title} {Time-dependent
  density functional theory with twist--averaged boundary conditions},}\ }\href
  {\doibase 10.1103/PhysRevC.93.054304} {\bibfield  {journal} {\bibinfo
  {journal} {Phys. Rev. C}\ }\textbf {\bibinfo {volume} {93}},\ \bibinfo
  {pages} {054304} (\bibinfo {year} {2016})}\BibitemShut {NoStop}%
\bibitem [{\citenamefont {Koonin}\ \emph {et~al.}(1977)\citenamefont {Koonin},
  \citenamefont {Davies}, \citenamefont {Maruhn-Rezwani}, \citenamefont
  {Feldmeier}, \citenamefont {Krieger},\ and\ \citenamefont
  {Negele}}]{koonin1977}%
  \BibitemOpen
  \bibfield  {author} {\bibinfo {author} {\bibfnamefont {S.~E.}\ \bibnamefont
  {Koonin}}, \bibinfo {author} {\bibfnamefont {K.~T.~R.}\ \bibnamefont
  {Davies}}, \bibinfo {author} {\bibfnamefont {V.}~\bibnamefont
  {Maruhn-Rezwani}}, \bibinfo {author} {\bibfnamefont {H.}~\bibnamefont
  {Feldmeier}}, \bibinfo {author} {\bibfnamefont {S.~J.}\ \bibnamefont
  {Krieger}}, \ and\ \bibinfo {author} {\bibfnamefont {J.~W.}\ \bibnamefont
  {Negele}},\ }\bibfield  {title} {\enquote {\bibinfo {title} {{T}ime-dependent
  {H}artree-{F}ock calculations for $^{16}${O} $+$ $^{16}${O} and $^{40}${C}a
  $+$ $^{40}${C}a reactions},}\ }\href {\doibase 10.1103/PhysRevC.15.1359}
  {\bibfield  {journal} {\bibinfo  {journal} {Phys. Rev. C}\ }\textbf {\bibinfo
  {volume} {15}},\ \bibinfo {pages} {1359--1374} (\bibinfo {year}
  {1977})}\BibitemShut {NoStop}%
\bibitem [{\citenamefont {Simenel}(2010)}]{simenel2010}%
  \BibitemOpen
  \bibfield  {author} {\bibinfo {author} {\bibfnamefont {C\'edric}\
  \bibnamefont {Simenel}},\ }\bibfield  {title} {\enquote {\bibinfo {title}
  {{P}article {T}ransfer {R}eactions with the {T}ime-{D}ependent
  {H}artree-{F}ock {T}heory {U}sing a {P}article {N}umber {P}rojection
  {T}echnique},}\ }\href {\doibase 10.1103/PhysRevLett.105.192701} {\bibfield
  {journal} {\bibinfo  {journal} {Phys. Rev. Lett.}\ }\textbf {\bibinfo
  {volume} {105}},\ \bibinfo {pages} {192701} (\bibinfo {year}
  {2010})}\BibitemShut {NoStop}%
\bibitem [{\citenamefont {{Kazuyuki Sekizawa}}\ and\ \citenamefont {{Kazuhiro
  Yabana}}(2013)}]{sekizawa2013}%
  \BibitemOpen
  \bibfield  {author} {\bibinfo {author} {\bibnamefont {{Kazuyuki Sekizawa}}}\
  and\ \bibinfo {author} {\bibnamefont {{Kazuhiro Yabana}}},\ }\bibfield
  {title} {\enquote {\bibinfo {title} {{T}ime-dependent {H}artree-{F}ock
  calculations for multinucleon transfer processes in
  $^{40,48}${C}a+$^{124}${S}n, $^{40}${C}a+$^{208}${P}b, and
  $^{58}${N}i+$^{208}${P}b reactions},}\ }\href {\doibase
  10.1103/PhysRevC.88.014614} {\bibfield  {journal} {\bibinfo  {journal} {Phys.
  Rev. C}\ }\textbf {\bibinfo {volume} {88}},\ \bibinfo {pages} {014614}
  (\bibinfo {year} {2013})}\BibitemShut {NoStop}%
\bibitem [{\citenamefont {Scamps}\ and\ \citenamefont
  {Lacroix}(2013)}]{scamps2013a}%
  \BibitemOpen
  \bibfield  {author} {\bibinfo {author} {\bibfnamefont {Guillaume}\
  \bibnamefont {Scamps}}\ and\ \bibinfo {author} {\bibfnamefont {Denis}\
  \bibnamefont {Lacroix}},\ }\bibfield  {title} {\enquote {\bibinfo {title}
  {{E}ffect of pairing on one- and two-nucleon transfer below the {C}oulomb
  barrier: {A} time-dependent microscopic description},}\ }\href {\doibase
  10.1103/PhysRevC.87.014605} {\bibfield  {journal} {\bibinfo  {journal} {Phys.
  Rev. C}\ }\textbf {\bibinfo {volume} {87}},\ \bibinfo {pages} {014605}
  (\bibinfo {year} {2013})}\BibitemShut {NoStop}%
\bibitem [{\citenamefont {Sekizawa}\ and\ \citenamefont
  {Yabana}(2014)}]{sekizawa2015}%
  \BibitemOpen
  \bibfield  {author} {\bibinfo {author} {\bibfnamefont {Kazuyuki}\
  \bibnamefont {Sekizawa}}\ and\ \bibinfo {author} {\bibfnamefont {Kazuhiro}\
  \bibnamefont {Yabana}},\ }\bibfield  {title} {\enquote {\bibinfo {title}
  {{P}article-number projection method in time-dependent {H}artree-{F}ock
  theory: {P}roperties of reaction products},}\ }\href {\doibase
  10.1103/PhysRevC.90.064614} {\bibfield  {journal} {\bibinfo  {journal} {Phys.
  Rev. C}\ }\textbf {\bibinfo {volume} {90}},\ \bibinfo {pages} {064614}
  (\bibinfo {year} {2014})}\BibitemShut {NoStop}%
\bibitem [{\citenamefont {{Sekizawa, Kazuyuki}}\ and\ \citenamefont {{Yabana,
  Kazuhiro}}(2015)}]{sekizawa2015b}%
  \BibitemOpen
  \bibfield  {author} {\bibinfo {author} {\bibnamefont {{Sekizawa, Kazuyuki}}}\
  and\ \bibinfo {author} {\bibnamefont {{Yabana, Kazuhiro}}},\ }\bibfield
  {title} {\enquote {\bibinfo {title} {Time-dependent {H}artree--{F}ock
  calculations for multi-nucleon transfer processes: {E}ffects of particle
  evaporation on production cross sections},}\ }\href {\doibase
  10.1051/epjconf/20158600043} {\bibfield  {journal} {\bibinfo  {journal}
  {{EPJ} {W}eb of {C}onf.}\ }\textbf {\bibinfo {volume} {86}},\ \bibinfo
  {pages} {00043} (\bibinfo {year} {2015})}\BibitemShut {NoStop}%
\bibitem [{\citenamefont {Dasso}\ \emph {et~al.}(1979)\citenamefont {Dasso},
  \citenamefont {Dossing},\ and\ \citenamefont {Pauli}}]{dasso1979}%
  \BibitemOpen
  \bibfield  {author} {\bibinfo {author} {\bibfnamefont {C.~H.}\ \bibnamefont
  {Dasso}}, \bibinfo {author} {\bibfnamefont {T.}~\bibnamefont {Dossing}}, \
  and\ \bibinfo {author} {\bibfnamefont {H.~C.}\ \bibnamefont {Pauli}},\
  }\bibfield  {title} {\enquote {\bibinfo {title} {{O}n the mass distribution
  in {T}ime-{D}ependent {H}artree-{F}ock calculations of heavy-ion
  collisions},}\ }\href {\doibase 10.1007/BF01409391} {\bibfield  {journal}
  {\bibinfo  {journal} {Z. Phys. A}\ }\textbf {\bibinfo {volume} {289}},\
  \bibinfo {pages} {395--398} (\bibinfo {year} {1979})}\BibitemShut {NoStop}%
\bibitem [{\citenamefont {Simenel}(2011)}]{simenel2011}%
  \BibitemOpen
  \bibfield  {author} {\bibinfo {author} {\bibfnamefont {C\'edric}\
  \bibnamefont {Simenel}},\ }\bibfield  {title} {\enquote {\bibinfo {title}
  {{P}article-{N}umber {F}luctuations and {C}orrelations in {T}ransfer
  {R}eactions {O}btained {U}sing the {B}alian-{V}\'en\'eroni {V}ariational
  {P}rinciple},}\ }\href {\doibase 10.1103/PhysRevLett.106.112502} {\bibfield
  {journal} {\bibinfo  {journal} {Phys. Rev. Lett.}\ }\textbf {\bibinfo
  {volume} {106}},\ \bibinfo {pages} {112502} (\bibinfo {year}
  {2011})}\BibitemShut {NoStop}%
\bibitem [{\citenamefont {Balian}\ and\ \citenamefont
  {V\'en\'eroni}(1992)}]{balian1992}%
  \BibitemOpen
  \bibfield  {author} {\bibinfo {author} {\bibfnamefont {R.}~\bibnamefont
  {Balian}}\ and\ \bibinfo {author} {\bibfnamefont {M.}~\bibnamefont
  {V\'en\'eroni}},\ }\bibfield  {title} {\enquote {\bibinfo {title}
  {Correlations and fluctuations in static and dynamic mean-field
  approaches},}\ }\href {\doibase 10.1016/0003-4916(92)90181-K} {\bibfield
  {journal} {\bibinfo  {journal} {Ann. Phys.}\ }\textbf {\bibinfo {volume}
  {216}},\ \bibinfo {pages} {351} (\bibinfo {year} {1992})}\BibitemShut
  {NoStop}%
\bibitem [{\citenamefont {Goutte}\ \emph {et~al.}(2005)\citenamefont {Goutte},
  \citenamefont {Berger}, \citenamefont {Casoli},\ and\ \citenamefont
  {Gogny}}]{goutte2005}%
  \BibitemOpen
  \bibfield  {author} {\bibinfo {author} {\bibfnamefont {H.}~\bibnamefont
  {Goutte}}, \bibinfo {author} {\bibfnamefont {J.~F.}\ \bibnamefont {Berger}},
  \bibinfo {author} {\bibfnamefont {P.}~\bibnamefont {Casoli}}, \ and\ \bibinfo
  {author} {\bibfnamefont {D.}~\bibnamefont {Gogny}},\ }\bibfield  {title}
  {\enquote {\bibinfo {title} {{M}icroscopic approach of fission dynamics
  applied to fragment kinetic energy and mass distributions in $^{238}${U}},}\
  }\href {\doibase 10.1103/PhysRevC.71.024316} {\bibfield  {journal} {\bibinfo
  {journal} {Phys. Rev. C}\ }\textbf {\bibinfo {volume} {71}},\ \bibinfo
  {pages} {024316} (\bibinfo {year} {2005})}\BibitemShut {NoStop}%
\bibitem [{\citenamefont {Ayik}(2008)}]{ayik2008}%
  \BibitemOpen
  \bibfield  {author} {\bibinfo {author} {\bibfnamefont {S.}~\bibnamefont
  {Ayik}},\ }\bibfield  {title} {\enquote {\bibinfo {title} {A stochastic
  mean-field approach for nuclear dynamics},}\ }\href {\doibase
  10.1016/j.physletb.2007.09.072} {\bibfield  {journal} {\bibinfo  {journal}
  {Phys. Lett. B}\ }\textbf {\bibinfo {volume} {658}},\ \bibinfo {pages} {174}
  (\bibinfo {year} {2008})}\BibitemShut {NoStop}%
\bibitem [{\citenamefont {Tohyama}(1985)}]{tohyama1985}%
  \BibitemOpen
  \bibfield  {author} {\bibinfo {author} {\bibfnamefont {M.}~\bibnamefont
  {Tohyama}},\ }\bibfield  {title} {\enquote {\bibinfo {title} {{T}wo-body
  collision effects on the low-{L} fusion window in $^{16}${O}+$^{16}${O}
  reactions},}\ }\href {\doibase 10.1016/0370-2693(85)91317-6} {\bibfield
  {journal} {\bibinfo  {journal} {Phys. Lett. B}\ }\textbf {\bibinfo {volume}
  {160}},\ \bibinfo {pages} {235--238} (\bibinfo {year} {1985})}\BibitemShut
  {NoStop}%
\bibitem [{\citenamefont {Tohyama}\ and\ \citenamefont
  {Umar}(2002)}]{tohyama2002a}%
  \BibitemOpen
  \bibfield  {author} {\bibinfo {author} {\bibfnamefont {M.}~\bibnamefont
  {Tohyama}}\ and\ \bibinfo {author} {\bibfnamefont {A.~S.}\ \bibnamefont
  {Umar}},\ }\bibfield  {title} {\enquote {\bibinfo {title} {{Q}uadrupole
  resonances in unstable oxygen isotopes in time-dependent density-matrix
  formalism},}\ }\href {\doibase 10.1016/S0370-2693(02)02885-X} {\bibfield
  {journal} {\bibinfo  {journal} {Phys. Lett. B}\ }\textbf {\bibinfo {volume}
  {549}},\ \bibinfo {pages} {72--78} (\bibinfo {year} {2002})}\BibitemShut
  {NoStop}%
\bibitem [{\citenamefont {{Marl\`ene Assi\'e}}\ and\ \citenamefont {{Denis
  Lacroix}}(2009)}]{assie2009}%
  \BibitemOpen
  \bibfield  {author} {\bibinfo {author} {\bibnamefont {{Marl\`ene Assi\'e}}}\
  and\ \bibinfo {author} {\bibnamefont {{Denis Lacroix}}},\ }\bibfield  {title}
  {\enquote {\bibinfo {title} {{P}robing {N}eutron {C}orrelations through
  {N}uclear {B}reakup},}\ }\href {\doibase 10.1103/PhysRevLett.102.202501}
  {\bibfield  {journal} {\bibinfo  {journal} {Phys. Rev. Lett.}\ }\textbf
  {\bibinfo {volume} {102}},\ \bibinfo {pages} {202501} (\bibinfo {year}
  {2009})}\BibitemShut {NoStop}%
\bibitem [{\citenamefont {Tohyama}\ and\ \citenamefont
  {Umar}(2016)}]{tohyama2016}%
  \BibitemOpen
  \bibfield  {author} {\bibinfo {author} {\bibfnamefont {M.}~\bibnamefont
  {Tohyama}}\ and\ \bibinfo {author} {\bibfnamefont {A.~S.}\ \bibnamefont
  {Umar}},\ }\bibfield  {title} {\enquote {\bibinfo {title} {{T}wo-body
  dissipation effects on the synthesis of superheavy elements},}\ }\href
  {\doibase 10.1103/PhysRevC.93.034607} {\bibfield  {journal} {\bibinfo
  {journal} {Phys. Rev. C}\ }\textbf {\bibinfo {volume} {93}},\ \bibinfo
  {pages} {034607} (\bibinfo {year} {2016})}\BibitemShut {NoStop}%
\bibitem [{\citenamefont {Ayik}\ \emph {et~al.}(2015)\citenamefont {Ayik},
  \citenamefont {Yilmaz},\ and\ \citenamefont {Yilmaz}}]{ayik2015a}%
  \BibitemOpen
  \bibfield  {author} {\bibinfo {author} {\bibfnamefont {S.}~\bibnamefont
  {Ayik}}, \bibinfo {author} {\bibfnamefont {B.}~\bibnamefont {Yilmaz}}, \ and\
  \bibinfo {author} {\bibfnamefont {O.}~\bibnamefont {Yilmaz}},\ }\bibfield
  {title} {\enquote {\bibinfo {title} {Multinucleon exchange in quasifission
  reactions},}\ }\href {\doibase 10.1103/physrevc.92.064615} {\bibfield
  {journal} {\bibinfo  {journal} {Phys. Rev. C}\ }\textbf {\bibinfo {volume}
  {92}},\ \bibinfo {pages} {064615} (\bibinfo {year} {2015})}\BibitemShut
  {NoStop}%
\bibitem [{\citenamefont {Lacroix}\ and\ \citenamefont
  {Ayik}(2014)}]{lacroix2014}%
  \BibitemOpen
  \bibfield  {author} {\bibinfo {author} {\bibfnamefont {Denis}\ \bibnamefont
  {Lacroix}}\ and\ \bibinfo {author} {\bibfnamefont {Sakir}\ \bibnamefont
  {Ayik}},\ }\bibfield  {title} {\enquote {\bibinfo {title} {{S}tochastic
  quantum dynamics beyond mean field},}\ }\href {\doibase
  10.1140/epja/i2014-14095-8} {\bibfield  {journal} {\bibinfo  {journal} {Eur.
  Phys. J. A}\ }\textbf {\bibinfo {volume} {50}},\ \bibinfo {pages} {95}
  (\bibinfo {year} {2014})}\BibitemShut {NoStop}%
\bibitem [{\citenamefont {Yilmaz}\ \emph {et~al.}(2014)\citenamefont {Yilmaz},
  \citenamefont {Ayik}, \citenamefont {Lacroix},\ and\ \citenamefont
  {Yilmaz}}]{yilmaz2014}%
  \BibitemOpen
  \bibfield  {author} {\bibinfo {author} {\bibfnamefont {B.}~\bibnamefont
  {Yilmaz}}, \bibinfo {author} {\bibfnamefont {S.}~\bibnamefont {Ayik}},
  \bibinfo {author} {\bibfnamefont {D.}~\bibnamefont {Lacroix}}, \ and\
  \bibinfo {author} {\bibfnamefont {O.}~\bibnamefont {Yilmaz}},\ }\bibfield
  {title} {\enquote {\bibinfo {title} {Nucleon exchange in heavy-ion collisions
  within a stochastic mean-field approach},}\ }\href {\doibase
  10.1103/physrevc.90.024613} {\bibfield  {journal} {\bibinfo  {journal} {Phys.
  Rev. C}\ }\textbf {\bibinfo {volume} {90}},\ \bibinfo {pages} {024613}
  (\bibinfo {year} {2014})}\BibitemShut {NoStop}%
\bibitem [{\citenamefont {Tanimura}\ \emph {et~al.}(2017)\citenamefont
  {Tanimura}, \citenamefont {Lacroix},\ and\ \citenamefont
  {Ayik}}]{tanimura2017}%
  \BibitemOpen
  \bibfield  {author} {\bibinfo {author} {\bibfnamefont {Yusuke}\ \bibnamefont
  {Tanimura}}, \bibinfo {author} {\bibfnamefont {Denis}\ \bibnamefont
  {Lacroix}}, \ and\ \bibinfo {author} {\bibfnamefont {Sakir}\ \bibnamefont
  {Ayik}},\ }\bibfield  {title} {\enquote {\bibinfo {title} {Microscopic
  {P}hase--{S}pace {E}xploration {M}odeling of $^{258}\mathrm{Fm}$
  {S}pontaneous {F}ission},}\ }\href {\doibase 10.1103/PhysRevLett.118.152501}
  {\bibfield  {journal} {\bibinfo  {journal} {Phys. Rev. Lett.}\ }\textbf
  {\bibinfo {volume} {118}},\ \bibinfo {pages} {152501} (\bibinfo {year}
  {2017})}\BibitemShut {NoStop}%
\bibitem [{\citenamefont {Gardiner}(1991)}]{gardiner1991}%
  \BibitemOpen
  \bibfield  {author} {\bibinfo {author} {\bibfnamefont {C.~W.}\ \bibnamefont
  {Gardiner}},\ }\href@noop {} {\emph {\bibinfo {title} {Quantum {N}oise}}}\
  (\bibinfo  {publisher} {Springer--Verlag},\ \bibinfo {address} {Berlin},\
  \bibinfo {year} {1991})\BibitemShut {NoStop}%
\bibitem [{\citenamefont {Weiss}(1999)}]{weiss1999}%
  \BibitemOpen
  \bibfield  {author} {\bibinfo {author} {\bibfnamefont {U.}~\bibnamefont
  {Weiss}},\ }\href@noop {} {\emph {\bibinfo {title} {Quantum {D}issipative
  {S}ystems}}},\ \bibinfo {edition} {2nd}\ ed.\ (\bibinfo  {publisher} {World
  {S}cientific},\ \bibinfo {address} {Singapore},\ \bibinfo {year}
  {1999})\BibitemShut {NoStop}%
\bibitem [{\citenamefont {Ayik}\ \emph {et~al.}(2016)\citenamefont {Ayik},
  \citenamefont {Yilmaz}, \citenamefont {Yilmaz},\ and\ \citenamefont
  {Umar}}]{ayik2016}%
  \BibitemOpen
  \bibfield  {author} {\bibinfo {author} {\bibfnamefont {S.}~\bibnamefont
  {Ayik}}, \bibinfo {author} {\bibfnamefont {O.}~\bibnamefont {Yilmaz}},
  \bibinfo {author} {\bibfnamefont {B.}~\bibnamefont {Yilmaz}}, \ and\ \bibinfo
  {author} {\bibfnamefont {A.~S.}\ \bibnamefont {Umar}},\ }\bibfield  {title}
  {\enquote {\bibinfo {title} {Quantal nucleon diffusion: {C}entral collisions
  of symmetric nuclei},}\ }\href {\doibase 10.1103/PhysRevC.94.044624}
  {\bibfield  {journal} {\bibinfo  {journal} {Phys. Rev. C}\ }\textbf {\bibinfo
  {volume} {94}},\ \bibinfo {pages} {044624} (\bibinfo {year}
  {2016})}\BibitemShut {NoStop}%
\bibitem [{\citenamefont {Umar}\ \emph {et~al.}(1991)\citenamefont {Umar},
  \citenamefont {Strayer}, \citenamefont {Wu}, \citenamefont {Dean},\ and\
  \citenamefont {G\"u\c{c}l\"u}}]{umar1991a}%
  \BibitemOpen
  \bibfield  {author} {\bibinfo {author} {\bibfnamefont {A.~S.}\ \bibnamefont
  {Umar}}, \bibinfo {author} {\bibfnamefont {M.~R.}\ \bibnamefont {Strayer}},
  \bibinfo {author} {\bibfnamefont {J.~S.}\ \bibnamefont {Wu}}, \bibinfo
  {author} {\bibfnamefont {D.~J.}\ \bibnamefont {Dean}}, \ and\ \bibinfo
  {author} {\bibfnamefont {M.~C.}\ \bibnamefont {G\"u\c{c}l\"u}},\ }\bibfield
  {title} {\enquote {\bibinfo {title} {{N}uclear {H}artree-{F}ock calculations
  with splines},}\ }\href {\doibase 10.1103/PhysRevC.44.2512} {\bibfield
  {journal} {\bibinfo  {journal} {Phys. Rev. C}\ }\textbf {\bibinfo {volume}
  {44}},\ \bibinfo {pages} {2512--2521} (\bibinfo {year} {1991})}\BibitemShut
  {NoStop}%
\bibitem [{\citenamefont {{Ka--Hae Kim}}\ \emph {et~al.}(1997)\citenamefont
  {{Ka--Hae Kim}}, \citenamefont {{Takaharu Otsuka}},\ and\ \citenamefont
  {{Paul Bonche}}}]{kim1997}%
  \BibitemOpen
  \bibfield  {author} {\bibinfo {author} {\bibnamefont {{Ka--Hae Kim}}},
  \bibinfo {author} {\bibnamefont {{Takaharu Otsuka}}}, \ and\ \bibinfo
  {author} {\bibnamefont {{Paul Bonche}}},\ }\bibfield  {title} {\enquote
  {\bibinfo {title} {{T}hree-dimensional {TDHF} calculations for reactions of
  unstable nuclei},}\ }\href {\doibase 10.1088/0954-3899/23/10/014} {\bibfield
  {journal} {\bibinfo  {journal} {J. Phys. G}\ }\textbf {\bibinfo {volume}
  {23}},\ \bibinfo {pages} {1267} (\bibinfo {year} {1997})}\BibitemShut
  {NoStop}%
\bibitem [{\citenamefont {Schr\"oder}\ \emph {et~al.}(1981)\citenamefont
  {Schr\"oder}, \citenamefont {Huizenga},\ and\ \citenamefont
  {Randrup}}]{schroder1981}%
  \BibitemOpen
  \bibfield  {author} {\bibinfo {author} {\bibfnamefont {W.~U.}\ \bibnamefont
  {Schr\"oder}}, \bibinfo {author} {\bibfnamefont {J.~R.}\ \bibnamefont
  {Huizenga}}, \ and\ \bibinfo {author} {\bibfnamefont {J.}~\bibnamefont
  {Randrup}},\ }\bibfield  {title} {\enquote {\bibinfo {title} {Correlated mass
  and charge transport induced by statistical nucleon exchange in damped
  nuclear reactions},}\ }\href {\doibase 10.1016/0370-2693(81)90924-2}
  {\bibfield  {journal} {\bibinfo  {journal} {Phys. Lett. B}\ }\textbf
  {\bibinfo {volume} {98}},\ \bibinfo {pages} {355--359} (\bibinfo {year}
  {1981})}\BibitemShut {NoStop}%
\bibitem [{\citenamefont {Merchant}\ and\ \citenamefont
  {N\"orenberg}(1981)}]{merchant1981}%
  \BibitemOpen
  \bibfield  {author} {\bibinfo {author} {\bibfnamefont {A.~C.}\ \bibnamefont
  {Merchant}}\ and\ \bibinfo {author} {\bibfnamefont {W.}~\bibnamefont
  {N\"orenberg}},\ }\bibfield  {title} {\enquote {\bibinfo {title} {Neutron and
  proton diffusion in heavy--ion collisions},}\ }\href {\doibase
  http://dx.doi.org/10.1016/0370-2693(81)90844-3} {\bibfield  {journal}
  {\bibinfo  {journal} {Phys. Lett. B}\ }\textbf {\bibinfo {volume} {104}},\
  \bibinfo {pages} {15--18} (\bibinfo {year} {1981})}\BibitemShut {NoStop}%
\bibitem [{\citenamefont {{Hannes Risken}}\ and\ \citenamefont {{Till
  Frank}}(1996)}]{risken1996}%
  \BibitemOpen
  \bibfield  {author} {\bibinfo {author} {\bibnamefont {{Hannes Risken}}}\ and\
  \bibinfo {author} {\bibnamefont {{Till Frank}}},\ }\href {\doibase
  10.1007/978-3-642-61544-3} {\emph {\bibinfo {title} {The {F}okker--{P}lanck
  {E}quation}}}\ (\bibinfo  {publisher} {Springer--Verlag},\ \bibinfo {address}
  {Berlin},\ \bibinfo {year} {1996})\BibitemShut {NoStop}%
\bibitem [{\citenamefont {N\"orenberg}(1981)}]{norenberg1981}%
  \BibitemOpen
  \bibfield  {author} {\bibinfo {author} {\bibfnamefont {W.}~\bibnamefont
  {N\"orenberg}},\ }\bibfield  {title} {\enquote {\bibinfo {title} {Memory
  effects in the energy dissipation for slow collective nuclear motion},}\
  }\href {\doibase 10.1016/0370-2693(81)90570-0} {\bibfield  {journal}
  {\bibinfo  {journal} {Phys. Lett. B}\ }\textbf {\bibinfo {volume} {104}},\
  \bibinfo {pages} {107--111} (\bibinfo {year} {1981})}\BibitemShut {NoStop}%
\bibitem [{\citenamefont {Randrup}(1979)}]{randrup1979}%
  \BibitemOpen
  \bibfield  {author} {\bibinfo {author} {\bibfnamefont {J.}~\bibnamefont
  {Randrup}},\ }\bibfield  {title} {\enquote {\bibinfo {title} {Theory of
  transfer-induced transport in nuclear collisions},}\ }\href {\doibase
  10.1016/0375-9474(79)90271-9} {\bibfield  {journal} {\bibinfo  {journal}
  {Nucl. Phys. A}\ }\textbf {\bibinfo {volume} {327}},\ \bibinfo {pages}
  {490--516} (\bibinfo {year} {1979})}\BibitemShut {NoStop}%
\bibitem [{\citenamefont {Merchant}\ and\ \citenamefont
  {N\"orenberg}(1982)}]{merchant1982}%
  \BibitemOpen
  \bibfield  {author} {\bibinfo {author} {\bibfnamefont {A.~C.}\ \bibnamefont
  {Merchant}}\ and\ \bibinfo {author} {\bibfnamefont {W.}~\bibnamefont
  {N\"orenberg}},\ }\bibfield  {title} {\enquote {\bibinfo {title} {Microscopic
  transport theory of heavy-ion collisions},}\ }\href {\doibase
  10.1007/bf01415880} {\bibfield  {journal} {\bibinfo  {journal} {Z. Phys. A}\
  }\textbf {\bibinfo {volume} {308}},\ \bibinfo {pages} {315--327} (\bibinfo
  {year} {1982})}\BibitemShut {NoStop}%
\end{thebibliography}%

\end{document}